\documentstyle[12pt]{article}

\advance \topmargin by -\headheight
\advance \topmargin by -\headsep
     
\evensidemargin \oddsidemargin
 \date{}    
\begin{document}
\newcommand{\sect}[1]{\setcounter{equation}{0}\section{#1}}
\renewcommand{\theequation}{\thesection.\arabic{equation}}

\topmargin -.6in
\def\nonu{\nonumber}
\def\rf#1{(\ref{eq:#1})}
\def\lab#1{\label{eq:#1}} 
\def\br{\begin{eqnarray}}
\def\er{\end{eqnarray}}
\def\be{\begin{equation}}
\def\ee{\end{equation}}
\def\0{\nonumber}
\def\lb{\lbrack}
\def\rb{\rbrack}
\def\({\left(}
\def\){\right)}
\def\v{\vert}
\def\bv{\bigm\vert}
\def\lskip{\vskip\baselineskip\vskip-\parskip\noindent}
\relax
\newcommand{\nit}{\noindent}
\newcommand{\ct}[1]{\cite{#1}}
\newcommand{\bi}[1]{\bibitem{#1}}
\def\a{\alpha}
\def\b{\beta}
\def\ca{{\cal A}}
\def\cm{{\cal M}}
\def\cn{{\cal N}}
\def\cf{{\cal F}}
\def\d{\delta} 
\def\D{\Delta}
\def\eps{\epsilon}
\def\g{\gamma}
\def\G{\Gamma}
\def\grad{\nabla}
\def\h{ {1\over 2}  }
\def\hc{\hat{c}}
\def\hd{\hat{d}}
\def\hg{\hat{g}}
\def\hp{ {+{1\over 2}}  }
\def\hm{ {-{1\over 2}}  }
\def\k{\kappa}
\def\l{\lambda}
\def\L{\Lambda}
\def\lg{\langle}
\def\m{\mu}
\def\n{\nu}
\def\o{\over}
\def\om{\omega}
\def\O{\Omega}
\def\p{\phi}
\def\pa{\partial}
\def\pr{\prime}
\def\ra{\rightarrow}
\def\rh{\rho}
\def\rg{\rangle}
\def\s{\sigma}
\def\t{\tau}
\def\th{\theta}
\def\ti{\tilde}
\def\wti{\widetilde}
\def\inte{\int dx }
\def\xb{\bar{x}}
\def\yb{\bar{y}}

\def\tr{\mathop{\rm tr}}
\def\Tr{\mathop{\rm Tr}}
\def\partder#1#2{{\partial #1\over\partial #2}}
\def\ds{{\cal D}_s}
\def\wtwo{{\wti W}_2}
\def\lie{{\cal G}}
\def\alie{{\widehat \lie}}
\def\dlie{{\cal G}^{\ast}}
\def\elie{{\widetilde \lie}}
\def\edlie{{\elie}^{\ast}}
\def\hlie{{\cal H}}
\def\wlie{{\widetilde \lie}}

\def\rlx{\relax\leavevmode}
\def\inbar{\vrule height1.5ex width.4pt depth0pt}
\def\IZ{\rlx\hbox{\sf Z\kern-.4em Z}}
\def\IR{\rlx\hbox{\rm I\kern-.18em R}}
\def\IC{\rlx\hbox{\,$\inbar\kern-.3em{\rm C}$}}
\def\one{\hbox{{1}\kern-.25em\hbox{l}}}

\def\PRL#1#2#3{{\sl Phys. Rev. Lett.} {\bf#1} (#2) #3}
\def\NPB#1#2#3{{\sl Nucl. Phys.} {\bf B#1} (#2) #3}
\def\NPBFS#1#2#3#4{{\sl Nucl. Phys.} {\bf B#2} [FS#1] (#3) #4}
\def\CMP#1#2#3{{\sl Commun. Math. Phys.} {\bf #1} (#2) #3}
\def\PRD#1#2#3{{\sl Phys. Rev.} {\bf D#1} (#2) #3}
\def\PLA#1#2#3{{\sl Phys. Lett.} {\bf #1A} (#2) #3}
\def\PLB#1#2#3{{\sl Phys. Lett.} {\bf #1B} (#2) #3}
\def\JMP#1#2#3{{\sl J. Math. Phys.} {\bf #1} (#2) #3}
\def\PTP#1#2#3{{\sl Prog. Theor. Phys.} {\bf #1} (#2) #3}
\def\SPTP#1#2#3{{\sl Suppl. Prog. Theor. Phys.} {\bf #1} (#2) #3}
\def\AoP#1#2#3{{\sl Ann. of Phys.} {\bf #1} (#2) #3}
\def\PNAS#1#2#3{{\sl Proc. Natl. Acad. Sci. USA} {\bf #1} (#2) #3}
\def\RMP#1#2#3{{\sl Rev. Mod. Phys.} {\bf #1} (#2) #3}
\def\PR#1#2#3{{\sl Phys. Reports} {\bf #1} (#2) #3}
\def\AoM#1#2#3{{\sl Ann. of Math.} {\bf #1} (#2) #3}
\def\UMN#1#2#3{{\sl Usp. Mat. Nauk} {\bf #1} (#2) #3}
\def\FAP#1#2#3{{\sl Funkt. Anal. Prilozheniya} {\bf #1} (#2) #3}
\def\FAaIA#1#2#3{{\sl Functional Analysis and Its Application} {\bf #1} (#2)
#3}
\def\BAMS#1#2#3{{\sl Bull. Am. Math. Soc.} {\bf #1} (#2) #3}
\def\TAMS#1#2#3{{\sl Trans. Am. Math. Soc.} {\bf #1} (#2) #3}
\def\InvM#1#2#3{{\sl Invent. Math.} {\bf #1} (#2) #3}
\def\LMP#1#2#3{{\sl Letters in Math. Phys.} {\bf #1} (#2) #3}
\def\IJMPA#1#2#3{{\sl Int. J. Mod. Phys.} {\bf A#1} (#2) #3}
\def\AdM#1#2#3{{\sl Advances in Math.} {\bf #1} (#2) #3}
\def\RMaP#1#2#3{{\sl Reports on Math. Phys.} {\bf #1} (#2) #3}
\def\IJM#1#2#3{{\sl Ill. J. Math.} {\bf #1} (#2) #3}
\def\APP#1#2#3{{\sl Acta Phys. Polon.} {\bf #1} (#2) #3}
\def\TMP#1#2#3{{\sl Theor. Mat. Phys.} {\bf #1} (#2) #3}
\def\JPA#1#2#3{{\sl J. Physics} {\bf A#1} (#2) #3}
\def\JSM#1#2#3{{\sl J. Soviet Math.} {\bf #1} (#2) #3}
\def\MPLA#1#2#3{{\sl Mod. Phys. Lett.} {\bf A#1} (#2) #3}
\def\JETP#1#2#3{{\sl Sov. Phys. JETP} {\bf #1} (#2) #3}
\def\JETPL#1#2#3{{\sl  Sov. Phys. JETP Lett.} {\bf #1} (#2) #3}
\def\PHSA#1#2#3{{\sl Physica} {\bf A#1} (#2) #3}
\def\PHSD#1#2#3{{\sl Physica} {\bf D#1} (#2) #3}

\begin{titlepage}
\vspace*{-2 cm}
\noindent
\begin{flushright}
\today 
\end{flushright}

\vskip 1 cm
\begin{center}
{\Large\bf Dyonic Integrable Models   } \vglue 1  true cm
{ J.F. Gomes}, E. P. Gueuvoghlanian,
 { G.M. Sotkov} and { A.H. Zimerman}\\

\vspace{1 cm}

{\footnotesize Instituto de F\'\i sica Te\'orica - IFT/UNESP\\
Rua Pamplona 145\\
01405-900, S\~ao Paulo - SP, Brazil}\\
jfg@ift.unesp.br, gueuvogh@ift.unesp.br, sotkov@ift.unesp.br, zimerman@ift.unesp.br\\

\vspace{1 cm}

\end{center}

\normalsize
\vskip 0.2cm

\begin{center}
{\large {\bf ABSTRACT}}\\
\end{center}
\noindent
A class of non abelian affine Toda models 
 arising from  the axial  gauged two-loop WZW model is presented. 
 Their zero curvature representation is  constructed in terms of a
graded Kac-Moody algebra.  It is shown that the discrete multivacua structure of the
potential together   with non abelian nature of the zero grade subalgebra
 allows soliton solutions with non trivial electric and topological charges. 
 The dressing transformation is employed to explicitly construct one and
 two soliton  solutions  and their bound states in terms of the tau functions.
A  discussion of the classical spectra of such solutions and the time delays  
 are given in detail.

  \noindent

\vglue 1 true cm

\end{titlepage}

\sect{Introduction}
The particle-like { nonperturbative } solutions of  the Yang-Mills-Higgs
(YMH) model and of the self-dual Yang-Mills (SDYM) equations - {\it monopoles,
dyons, instantons, etc }, as well as their { extended } 
solutions -{\it domain walls} (DW) and  {\it strings }- 
are known to have many common features with the
soliton solutions of certain 1- and 2-D integrable models (IM). 
Each of these solutions corresponds to specific boundary conditions (b.c.'s) 
imposed on the 4-D fields,
 $A_{\mu}^{a}$ and $ \phi^{a}$ at space infinities, that in general break  
 translational invariance, Lorentz rotations and the gauge
 symmetry.  As a consequence the 4-D SDYM (or YMH- Bogomolny) equations 
 are reduced to lower dimensional { integrable} equations. 
 The most representative example is the spherical symmetric 1-monopole of 
 the $SU(r+1)$-YMH model described in terms of specific solutions
 of 1-D $SU(r+1)$- abelian Toda theory \cite{ganoulis}. 
 
  Recently, solitons 
 of 2-D $N=2$ supersymmetric $CP^r$ model ( and its mirror dual $N=2$ $SU(r+1)$
abelian affine Toda model )have been identified as BPS {\it domain walls} of 
4-D $N=1$ supersymmetric $SU(r+1)$
YM theory \cite{shiff1},\cite{hh},\cite{witt1}.  Since 
large $r$ limits of $N=1$ SUSY YM and of the ordinary QCD are
believed to coincide \cite{witt2}, one expects that domain walls of 
non SUSY YMH (for large $r$ and maximal breaking 
$SU(r+1)\rightarrow U(1)^r$) to be related to solitons of 2-D 
$SU(r+1)$ abelian affine Toda model.  The  known exact
solitonic results of the 2-D IMs can than be used to describe 
nonperturbative properties of the domain walls \cite{shiff1}, \cite{hh}.  
The domain wall's tensions, for
instance are  proportional  to the   
soliton masses. DW's junctions are related to the multisolitons, 
breathers ans excited solitons, etc.

In (super)string   aproach to 4-D YMH theory, the DW's appear 
as specific $D2$-branes where QCD string can end \cite{witt1}. 
 An interesting  ``soliton solution '' of 10-D (or 26-D) low energy
effective field theory requires together with simple DW's, for instance, $U(1) $ charged DW's,
$n$- coincident $D2$-branes, nonstatic (but stationary ) DW's, etc. 
 The explicit description of such DW's in terms of 2-D solitons
addresses {\it the problem} of constructing new families of IMs of dyonic 
type admiting $U(1)$ charged topological solitons, global (or local )
$U(1)$ Noether symmetry, CPT breaking $\theta$-terms (see for instance \cite{t})  The 
abelian affine Toda theories are examples of IM's satisfying none of these
requirements.  It is natural to seek for such dyonic IM's within the 
class of non-abelian affine Toda models \cite{lez-sav},\cite{mira}.
Integrable models with $U(1)$ (or
more general $U(1)^{r}$) symmetry such as the Lund-Regge (LR) (or complex
Sine-Gordon, CSG) model \cite{lundregge} and its homogeneous space SG generalizations
\cite{miramontes} appear to be of purely {\it electric } type: their solitons
are   $U(1)^{r}$ {\it charged} but nontopological.  The simplest IM of {\it
dyonic } type with Lagrangean 
\br
{\cal L} &=& {1\o 2} \sum_{i,j=2}^{r} K_{ij} \pa^{\mu} \varphi_i \pa _{\mu} \varphi_j +
{{e^{-\b \varphi_2}}} {{\( \pa ^{\mu}\psi \pa _{\mu }\chi +
\eps^{\mu \nu } \pa _{\mu} \psi \pa _{\nu} \chi \)}
 \o {1+ \b^2 {{n+1}\o {2n}}\psi \chi e^{-\b \varphi_2} }} \nonu \\
 &-& {{\mu ^2 }\o {\b^2}} \( \sum_{i=2}^{r} e^{-\b K_{ij} \varphi_j} + e^{\b
 (\varphi_{r} + \varphi_{2})}(1+ \b^2 \psi \chi e^{-\b \varphi_2})-r \)
 \label{1.1}
 \er
 $K_{ij} = 2 \d_{ij} - \d_{i, j-1} -\d_{i, j+1}, \;\; i,j = 2, \cdots r$,
 have been studied in our recent paper \cite{backlund}.  
   For imaginary coupling $\b
 \rightarrow i \b_0$ its potential  has $r-1$ distinct vacua, 
 $Z_2 \otimes Z_{r-1}$ symmetry and as a consequence the model (\ref{1.1})
 admits, together with the {\it neutral topological } solitons
  (of the $A_{r-1}^{(1)}$ abelian affine Toda type ), the $U(1)$ charged
 topological 1-solitons with spectra quite similar to the YMH 
 dyons \cite{backlund}.  These dyonic type of solitons share many of 
 the properties of the charged DWs in the $N=1$ SYM for nonmaximal
 breaking $SU(r+1) \rightarrow SU(2)\otimes U(1)^{r-1}$.

The present paper is devoted to the systematic construction of IM of {\it dyonic}
type and of their multisoliton and breather solutions.
  The method used in the derivation of their action represents a functional
integral version of the Leznov-Saveliev algebraic approach \cite{lez-sav} to the $\hat
\lie _r$ non-abelian (NA) affine Toda IM.  It is explained in Sect. 2 that a large
class of 2-D IM can be defined as
 {\it axial } (or vector ) gauged {\it two loop}
WZW models $  \hat {H}_- \backslash \hat {G}_r / \hat {H}_+$ \cite{Aratyn}
 by extending the methods
for constructing conformal $\s$-models  and conformal NA-Toda \cite{plb} to
include all integrable (non conformal ) perturbations.  As it is well known each
affine (Kac-Moody) algebra $\hat {G}_r$ gives rise to many different IM's depending
on the choice of specific graded structure defined in terms of a grading operator 
denoted by $Q$  and by the  constant grade $\pm 1$ elements
$\eps_{\pm}$ (or $\pm p$, for higher grading model).

The zero grade (factor) group  $\G_0 =  \hat {H}_- \backslash \hat {G}_r / \hat {H}_+$ is
parametrized by the (physical ) fields (say $\varphi_i, \psi, \chi $ ) present in the
action \footnote{ Note that the nonconformal 
affine NA Toda models are constructed in terms of the centerless Kac-Moody algebra, 
i.e. $\hat c = 0$}.
  The elements of the zero grade subalgebras $\lie_0^0 \subset \lie_0$ such that 
$ [\lie^0_0, \eps_{\pm} ] =0 $, 
generates the (Noether) symmetries of the corresponding IM. They are by
construction, local symmetries and when gauged away, give rise to the so called
singular NA affine Toda with $\lie_0^0$ as  the {\it global} symmetry group.  
The soliton charges
are shown to depend upon  $\lie_0^0$ and  upon the specific form of $\eps_{\pm}$ 
 which defines the  potential  of our IM $V= Tr \( B^{-1} \eps_+ B
\eps_- \), \;\; (B \in G_0 / G_0^0)$ and on the boundary condition (b.c.'s ) imposed upon the
fields.  The model of {\it magnetic } type has no Noether charges $\{\lie_0^0 \} = \emptyset $ 
and $\eps_{\pm}$ are chosen in such way that provides
nontrivial zeros of the potential $V$. They have {\it maximal } number of 
{\it magnetic } (i.e. topological ) charges $m_i: \Lambda = {{2\pi }\o
{\b}}\sum_{i=1}^r m_i \a_i^{v}$.  The {\it electric } type IM are characterized by
the {\it maximal } number of Noether charges $\lie_0^0 \equiv U(1)^{r}$ that
restrict $\eps_{\pm} = \sum_{i=1}^r \mu_i h_i^{(\pm 1)}, \; (Q= \hat d)$,
 (homogeneous gradation) and for 
the standard b.c. (all fields vanish at $\pm \infty $ ) leads to potentials with
trivial magnetic charges.  The main characteristic of the IM of {\it dyonic } type
is that only a part of the Cartan subalgebra ${\cal H}_r = \oplus_1^r U(1)$, (i.e.
$\lie_0^0 = U(1)^{s}, \; 0\leq s \leq r$) generates Noether symmetries (i.e.
the number of $U(1)$ charges is  $0\leq s \leq r$) and the  rest $r-s$ can be
converted into magnetic charges $m_a, \;  a=s+1, \cdots , r$ if  $\eps_{\pm }$
are appropriately choosen.  The generic type $U(1)$ dyonic IM (i.e. $s=1$) is
defined by  \be
Q = \tilde h \hat d +
\sum_{i\neq a}^{r}{{2 \l_i \cdot H^{(0)}} \o {\a_i^2}}, \quad 
\eps_{\pm} = \sum_{i\neq a} \mu_i E_{\pm \a_i}^{(0)} + 
\sum_{l} \mu_{0l}E_{\mp \b_{0l}}^{(\pm 1)}
\label{1.4}
\ee
$\lie_0^0 = \{ \l_{a} \cdot H^{(0)}, 1\leq a \leq r,\}$  $a$  fixed, $\a_i $ are the simple
roots of $\hat G_r$ and $\b_{0l}$ are specific composite roots.  The exact algebraic
criterium for the construction of $U(1)$ dyonic IMs of axial type is given in Sect. 2. 
 The condition for absence (or presence ) of CPT-breaking terms as well as 
 the derivation of the CPT-invariant dyonic  IMs of the vector type ( the T-duals of (\ref{1.1}))
  are presented in \cite{backlund} and \cite{wigner99} . 
  The IM
(\ref{1.1}) corresponds to $\hat G_r = A_r^{(1)}, \;\; a = 1 $ (i.e. $\lie_0^0 =
\l_1 \cdot H^{(0)}$), $\b_0 = \a_2 + \a_3  +\cdots + \a_r$, $\tilde h = r$ and $\m_i = \mu_{0l} = \mu$.  

We further choose to study the IM (\ref{1.1}) as  representative of the family
of a single $U(1)$ charge IM described in Sect. 2.  The remaining part of the
paper (Sects. 3-8) contains the explicit constructions of all 1- and 2-
soliton solutions and their bound states by employing the method of dressing
transformation (or `` vertex operators'', or ``soliton specialization''
``$\tau$-function''). The complete list of ``irreducible''  particle-like
solutions  contains
together with the $U(1)$ {\it charged} topological 1-soliton  also {\it neutral } topological 1-solitons 
 of species $a$ \cite{liao}, and their {\it bound states}:
\begin{itemize}
\item ``2-neutral vertex''  solution,  that coincides
with $A_{r-1}^{(1)}$ abelian affine Toda  bound states  \cite{mcghee}.

\item ``3-vertex ''  describing the one charged 1-soliton and one
neutral 1-soliton bound state

\item ``4-vertex''  representing a bound state of two charged 1-(anti) solitons.
\end{itemize} 

An important feature of the neutral and charged (multi) solitons and bound states of IMs (\ref{1.1}) 
is the so called vacuum degeneration.  It consists in the
 existence of  soliton solutions with same energy, momentun, mass,  
electric charge, but with different number
(moduli ) parameters and different form of the solutions (i.e. $\tau $-functions ).  They are constructed by appropriated modification 
of  the defining constant 
element $g$, characteristic of the dressing method (see Sect. 4.3 and 5).

The scattering phase shifts (time delays )  are derived in Sect.6 from the
explicit form of the various 2-soliton - antisoliton solutions.

\sect{Construction of the Model}
The generic NA Toda models  are classified
 according to a $\lie_0 \subset \lie$ embedding  induced
by the grading operator $Q$ decomposing an finite or infinite Lie algebra 
$\lie = \oplus _{i} \lie _i $ where $
[Q,\lie_{i}]=i\lie_{i}$ and $ [\lie_{i},\lie_{j}]\subset \lie_{i+j}$.  A group
element $g$ can then be written in terms of the Gauss decomposition as 
\be
g= NBM
\label{1}
\ee
where $N=\exp \lie_< $, $B=\exp \lie_{0} $ and
$M=\exp \lie_> $.  The physical fields lie in the zero grade subgroup $B$ 
and the
models we seek correspond to the coset $H_- \backslash G/H_+ $, for $H_{\pm} $ generated by
positive/negative  grade operators.

For consistency with the hamiltonian reduction formalism, the phase space of
the G-invariant WZNW model is  reduced by specifying the constant
generators $\eps_{\pm}$ of grade $\pm 1$.  In order to derive 
 an action for $B  $,  invariant under 
\begin{eqnarray}
g\longrightarrow g^{\prime}=\alpha_{-}g\alpha_{+},
\label{2}
\end{eqnarray}
where $\a_{\pm}(z, \bar z)$ lie in the positive/negative grade subgroup
 we have to introduce a set of  auxiliary
gauge fields $A \in \lie _{<} $ and $\bar A \in \lie _{>}$ transforming as 
\begin{eqnarray}
A\longrightarrow A^{\prime}=\alpha_{-}A\alpha_{-}^{-1}
+\alpha_{-}\partial \alpha_{-}^{-1},
\quad \quad 
\bar{A}\longrightarrow \bar{A}^{\prime}=\alpha_{+}^{-1}\bar{A}\alpha_{+}
+\bar{\partial}\alpha_{+}^{-1}\alpha_{+},
\label{3}
\end{eqnarray}
where $z = {1\o 2} (t+x), \bar z = {1\o 2} (t-x)$.
 The resulting action is the $G/H (= H_- \backslash G/H_+ )$
 gauged WZNW    
\begin{eqnarray}
S_{G/H}(g,A,\bar{A})&=&S_{WZNW}(g)
\nonumber
\\
&-&\frac{k}{2\pi}\int d^2x Tr\( A(\bar{\partial}gg^{-1}-\epsilon_{+})
+\bar{A}(g^{-1}\partial g-\epsilon_{-})+Ag\bar{A}g^{-1}\) .
\nonumber
\end{eqnarray}
Since the action $S_{G/H}$ is $H$-invariant,
 we may choose $\alpha_{-}=N_{}^{-1}$
and $\alpha_{+}=M_{}^{-1}$. From the orthogonality  of the graded 
subpaces, i.e. $Tr( \lie _i\lie _j ) =0, i+j \neq 0$, we find
\begin{eqnarray}
S_{G/H}(g,A,\bar{A})&=&S_{G/H}(B,A^{\prime},\bar{A}^{\prime})
\nonumber
\\
&=&S_{WZNW}(B)-\frac{k}{2\pi}
\int d^2x Tr[-A^{\prime}\epsilon_{+}-\bar{A}^{\prime}\epsilon_{-}
+A^{\prime}B\bar{A}^{\prime}B^{-1}],
\label{14}
\end{eqnarray}
where 
\begin{eqnarray}
S_{WZNW}=- \frac{k}{4\pi }\int d^2xTr(g^{-1}\partial gg^{-1}\bar{\partial }g)
+\frac{k}{24\pi }\int_{D}\epsilon^{ijk}
Tr(g^{-1}\partial_{i}gg^{-1}\partial_{j}gg^{-1}\partial_{k}g)d^3x,
\label{3a}
\end{eqnarray}
and the topological term denotes a surface integral  over a ball $D$
identified as  space-time.

Action (\ref{14}) describes the non singular Toda models among which we find the
Conformal and the Affine abelian Toda models where 
$Q=\sum_{i=1}^{r}\frac{2\lambda_{i}\cdot H}{\alpha_{i}^{2}}, \quad 
 \epsilon_{\pm}=\sum_{i=1}^{r} \mu_{\pm i}E_{\pm \alpha_{i}}$ and 
$Q= h d + \sum_{i=1}^{r}\frac{2\lambda_{i}\cdot H^{(0)}}{\alpha_{i}^{2}}, \quad 
\epsilon_{\pm}=\sum_{i=1}^{r} \mu_{\pm i}E_{\pm \alpha_{i}}^{(0)} + 
\mu_{0}E_{\mp \psi }^{(\pm 1)}$
respectively,  $\psi $ denote the highest root,  $\lambda_i$ are the 
fundamental weights  and $h$ the coxeter number of $\lie $. 

Performing the integration  over the auxiliary fields $A$ and $\bar A$, 
the functional integral 
\be
 Z_{\pm}=\int DAD\bar{A}\exp (F_{\pm}),
\label{fpm}
\ee 
 where 
 \be
F_{\pm} = {-{k\o {2\pi}}}\int \(Tr
 (A - B {\eps_-} B^{-1})B(\bar A - B^{-1} {\eps_+} B) B^{-1}\)d^2x
\label{fmm}
\ee
yields the effective action
\be
S = S_{WZNW} (B) + {{k\o {2\pi}}} \int Tr \( \eps_+ B  \eps_- B^{-1}\)d^2x
\label{spm}
\ee
The action (\ref{spm}) describes integrable perturbations of the $\lie_0$-WZNW model. 
 Those perturbations are classified in
terms of the possible constant grade $\pm 1$ operators $\eps _{\pm}$.

 More interesting cases 
arises in
connection with non abelian embeddings $\lie_0 \subset \lie $.  In particular, if we 
supress one
of the fundamental weights from $Q$, the zero grade subspace $\lie_0$,
acquires a nonabelian 
structure
$sl(2)\otimes u(1)^{rank \lie -1}$.  Let us consider for instance 
$Q= h^{\pr} d + \sum_{i\neq a}^{r}\frac{2\lambda_{i}\cdot H}{\alpha_{i}^{2}}$, 
where $h^{\pr} =0$ or
$h^{\pr} \neq 0$ \footnote{ For the Kac-Moody case we are suppressing the
index $(0)$ in defining the Cartan subalgebra of $\lie $} corresponding to  the
Conformal or Affine  nonabelian (NA) Toda  respectively. The absence of
$\lambda_a$ in $Q$  prevents the contribution of the simple root step operator
$E_{\a_a}^{(0)}$ in constructing $\eps_+$. It in fact, allows for  reducing
the phase space  even further.  This fact can be understood by enforcing the 
nonlocal constraint $J_{Y \cdot H} = \bar J_{Y \cdot H} = 0$
where $Y$ is such that $[Y\cdot H , \eps_{\pm}] = 0$ and 
$J=g^{-1}\partial g$ and $\bar{J}=-\bar{\partial}gg^{-1}$.  Those generators 
of $\lie_0$ commuting with
$\eps_{\pm}$ define a subalgebra $\lie_0^0 \subset \lie_0 $.
 Such subsidiary constraint is incorporated into the action by
  requiring symmetry under \cite{plb} 
\begin{eqnarray}
g\longrightarrow g^{\prime}=\alpha_{0}g\alpha_{0}'
\label{5}
\end{eqnarray} 
where we shall consider   $\a^{\pr}_{0}
 =\alpha_{0}(z, \bar z) \in \lie_0^0 $, i.e., {\it axial symmetry} (the 
 {\it vector }
 gauging is obtained by choosing $\a^{\pr}_{0}
 ={\alpha_{0}}^{-1}(z, \bar z) \in \lie_0^0 $, see for instance the second ref. in \cite{wigner99}). 
Auxiliary gauge fields $A_0 = a_0 Y\cdot H$ and $\bar A_0= 
\bar a_0 Y\cdot H\in  \lie_{0}^{0}$  are 
introduced to construct  an invariant action under transformations  (\ref{5}) 
\br 
S(B,{A}_{0},\bar{A}_{0} ) &=& S(g_0^f,{A^{\pr}}_{0},\bar{A^{\pr}}_{0} )  
 = S_{WZNW}(B)+ 
 {{k\o {2\pi}}} \int Tr \( \eps_+ B  \eps_- B^{-1}\) d^2x\nonu \\   
  &-&{{k\o {2\pi}}}\int Tr\(  A_{0}\bar{\partial}B
B^{-1} + \bar{A}_{0}B^{-1}\partial B
+ A_{0}B\bar{A}_{0}B^{-1} + A_{0}\bar{A}_{0} \)d^2x \nonu \\
\label{aa}
\er
where the auxiliary fields transform as
\begin{eqnarray}
A_{0}\longrightarrow A_{0}^{\prime}=A_{0}-\alpha_{0}^{-1}\partial \alpha_{0},
\quad \quad 
\bar{A}_{0}\longrightarrow \bar{A}_{0}^{\prime}=\bar{A}_{0}
- \bar{\partial}\alpha_{0}^{\pr}(\alpha_{0}^{\pr})^{-1}.
\nonu 
\end{eqnarray}
 Such residual gauge symmetry allows us to eliminate an  extra field
 associated to $Y\cdot H$.
  Notice that the physical fields
 $g_0^f$ lie in the coset $\lie_0 /{\lie_0^0} = ({{sl(2)\otimes
u(1)^{rank \lie -1}})/u(1)}$ of dimension $rank
\lie +1$ and are classified according to the 
gradation $Q$.  It therefore follows that $S(B,A_0,\bar{A_0})=
 S(g_0^f,A_0^{\pr},\bar{A_0}^{\pr})$.

In \cite{ime} a detailed study of the gauged WZNW construction 
for  finite dimensional Lie algebras leading to Conformal NA Toda
models was presented.  The study of its symmetries was given in refs. \cite{plb}. 
 Here we generalize the construction of ref. \cite{ime} to
infinite dimensional Lie algebras leading to NA Affine 
Toda models characterized 
by the broken
conformal symmetry and by the  presence of
solitons. 
 
Consider the Kac-Moody
algebra ${\widehat \lie }$
\br
[T_m^a ,T_n^b]  =  f^{abc} T^c_{m+n} + {\hat c}m \d_{m+n} \d^{ab} \nonu 
\er
\br
[{\hat d} , T^a_n] = nT^a_n ;\quad [{\hat c}, T^a_n] = [{\hat c},{\hat d} ] = 0
\label{km}
\er

The NA Toda models we shall be constructing are associated to 
gradations of the type
$Q_a(h^{\pr}) = h^{\pr}_a d + 
\sum_{i\neq a}^{r}\frac{2\lambda_{i}\cdot H^{(0)}}{\alpha_{i}^{2}}$, 
where $h^{\pr}_a$ is chosen
such that the gradation,
 $Q_a(h^{\pr})$, acting on
infinite dimensional Lie algebra $ \hat \lie$ ensures that the zero grade 
subgroup $\lie_0$ coincides with its
counterpart obtained with $Q_a(h^{\pr}=0)$ 
 acting on the Lie algebra $\lie $ of finite dimension apart from two  commuting 
 generators $\hat {c}$ and $\hat {d}$.  Since they commute with $\lie_0$,
  the kinetic part
 decouples such that the conformal and the affine singular NA-Toda  models 
 differ only by the
 potential term characterized by $\hat \eps_{\pm}$. 

The integration over the auxiliary gauge fields $A$ and $\bar A$ requires 
explicit
parametrization of $B$. 
\begin{eqnarray}
B=\exp (\tilde {\chi} E_{-\alpha_{a}}^{(0)})
 \exp (   R \sum_{i=1}^{r}{{Y_i}} {{H_i^{(0)}}}+\Phi (H)+ \nu \hat {c} +
  \eta \hat {d})\exp (\tilde {\psi} E_{\alpha_{a}}^{(0)})
 \label{63}
 \end{eqnarray}
where $ \Phi (H) 
=\sum_{j=1}^{r}\sum_{i=2}^{r}\varphi_{i}{{X}}_i^j {{H_j}^{(0)}}$,
 where $\sum_{j=1}^r{{Y_j}}  {{X^j_i}} =0, i=2, \cdots ,r$.
After gauging away the nonlocal field $R$, the factor group element becomes    
\be
g_0^f=\exp (\chi E_{-\alpha_{a}}^{(0)})
 \exp (   \Phi (H)+ \nu \hat {c} + \eta \hat {d})\exp (\psi E_{\alpha_{a}}^{(0)})
 \label{63a}
\ee
where $\chi = \tilde {\chi}e^{{1\o 2}(Y\cdot \a_a)R}, \quad 
\psi = \tilde {\psi}e^{{1\o 2}(Y\cdot \a_a)R}$.
We therefore get for the zero grade component
\br
F_0 &=&{-{k\o {2\pi}}}\int Tr\(  A_{0}\bar{\partial}g_0^f
(g_0^f)^{-1} + \bar{A}_{0}(g_0^f)^{-1}\partial g_0^f
+ A_{0} g_0^f\bar{A}_{0}(g_0^f)^{-1} + A_{0}\bar{A}_{0} \)d^2x \nonu \\
&=&{-{k\o {2\pi}}}\int \( a_0 \bar a_0 2Y^2\Delta -  2({{\a_a \cdot Y}\o
{\a_a^2}})(\bar a_0\psi \pa \chi + a_0 \chi \bar \pa \psi )e^{\Phi (\a_a)}\)
d^2x
\label{del}
\er
where $\Delta = 1 + {{(Y \cdot \a_a )^2}\o {2Y^2}}\psi \chi e^{\Phi (\a_a )}$,
$[\Phi (H), E_{\alpha_{a}}^{(0)}] = \Phi (\a_a)E_{\alpha_{a}}^{(0)}$. 

The effective action is obtained by integrating over the auxiliary 
fields $A_0, \bar A_0$, 
\be
Z_0 = \int DA_{0}D\bar{A}_{0}\exp (F_{0}) 
\ee  
 The total action (\ref{aa}) is therefore given as
 \be
 S= -{k \o {4\pi}}\int \( Tr (\pa \Phi(H)\bar \pa \Phi(H)) + 
 {{2\bar \pa \psi \pa \chi }\o \Delta
 }e^{\Phi(\a_a)} + \pa \eta \bar \pa \nu +
 \pa \nu \bar \pa \eta 
    - 2 Tr (\hat {\eps_+} g_0^f\hat {\eps_-} (g_0^f)^{-1}) \)d^2x
 \label{action}
 \ee
Note that the second term in (\ref{action}) contains both symmetric and
antisymmetric parts:
\begin{eqnarray}
\frac{e^{\Phi(\a_a)} } {\Delta}\bar{\partial}\psi \partial \chi
=\frac{ e^{\Phi(\a_a)} }
{\Delta}(g^{\mu \nu}\partial_{\mu}\psi\partial_{\nu}\chi
+\epsilon^{\mu \nu}\partial_{\mu}\psi \partial_{\nu}\chi ),
\label{axx1}
\end{eqnarray}
where $g_{\mu \nu}$ is the 2-D metric of signature $ g_{\mu
\nu}= diag (1,-1)$, $\eps^{\mu \nu}$ is the totally antisymmetric tensor $\eps^{01} = 1$.
 For $n=1$ ($\lie \equiv A_{1}$, $\Phi (\a_1)$ is zero) the
antisymmetric term is a total derivative:
\begin{eqnarray}
\epsilon^{\mu \nu}\frac{\partial_{\mu}\psi \partial_{\nu}\chi}{1+\psi \chi}
=\frac{1}{2}\epsilon^{\mu \nu}\partial_{\mu}
\left( \ln \left\{ 1+\psi \chi \right\}
\partial_{\nu}\ln{\frac{\chi}{\psi}}\right),
\label{axx2}
\end{eqnarray}
and it can be neglected.  This $A_{1}$-NA-Toda model (in the conformal case), 
is known to describe the
2-D black hole solution for (2-D) string theory \cite{Witten1}.
The
$\lie $-NA conformal Toda model can be used in the
description of specific (n+1)-dimensional black string theories 
\cite{gervais-saveliev},
 with  (n-1)-flat and
2-non flat directions ($g^{\mu
\nu}G_{ab}(X)\partial_{\mu}X^{a}\partial_{\nu}X^{b}$, $X^{a}=(\psi ,\chi
,\varphi_{i})$), containing axions ($\epsilon^{\mu
\nu}B_{ab}(X)\partial_{\mu}X^{a}\partial_{\nu}X^{b}$) and tachions
($\exp \left\{ -k_{ij}\varphi_{j}\right\} $), as well.  The affine 
$ A_{1}$-NA Toda theory with $\eps _{\pm} = H^{(\pm )}$
corresponds to the Lund-Regge model describing charged solitons 
\cite{hollowood}.

It is clear that the presence of the $ e^{\Phi(\a_a)}$ in (\ref{action})
is responsible for the
 antisymmetric  tensor
generating the axionic terms. 
On the other hand, notice that $\Phi(\a_a)$ depends upon the subsidiary nonlocal
constraint $J_{Y \cdot H} = \bar J_{Y \cdot H} = 0$ and hence upon the choice of
the vector {{Y}}.  It is defined to be orthogonal to all roots contained in
$\eps_{\pm}$.  A Lie algebraic condition for the absence of axionic terms was found in \cite{ime} and
has provided a construction of a family of torsionless NA Toda models in 
\cite{wigner99}.

The action (\ref{action}) is invariant under the global $U(1)$ transformation 
\begin{equation}
{\psi }\rightarrow e^{i\epsilon }{\psi }, \quad \quad 
{\chi }\rightarrow e^{-i\epsilon }{\chi }
 \label{13.84}
\end{equation}
The corresponding Noether current is 
\begin{equation}
J^{\mu }=-\frac{ik}{2\pi }\frac{e^{\Phi(\a_a)}}{\Delta }\{{\psi 
}\left( g^{\nu \mu }\partial _{\nu }{\chi }-\epsilon ^{\nu \mu
}\partial _{\nu }{\chi }\right) -{\chi }\left( g^{\nu
\mu }\partial _{\nu }{\psi }+\epsilon ^{\nu \mu }\partial _{\nu }
{\psi }\right) \}
\end{equation}
and the electric charge is given in terms of the nonlocal field $R$ defined 
below in
(\ref{bh}) as
\begin{equation}
Q_{}=\int J_o dx= -\frac{ik}{\pi }\left( \frac{r}{r+1}\right) \left[
R(\infty )-R(-\infty )\right].  \label{13.87}
\end{equation}

Apart from the global $U(1)$   symmetry (\ref{13.84}) there is a discrete 
set of field transformations leaving the action (\ref{action}) unchanged.
Such transformations  (for imaginary $\b $, $\b \rightarrow i \b_0$) 
 give rise to multivacua configuration and 
hence to nontrivial topological charges
\br
 Q_j &=& \int_{-\infty}^{+\infty} J^0_{j} dx,  \quad J_j^{\mu} = 
  {{1\o {2\pi i}}}\eps^{\mu \nu } \pa_{\nu} \varphi_j, \quad j=2,
 \cdots r \nonu \\
 Q_{\theta} &=& \int_{-\infty}^{+\infty} J_{\theta}^{0} dx    
  \quad J_{\theta}^{\mu} = -i {{1}\o { 2\b^2}} \eps^{\mu \nu } \pa_{\nu} 
 ln \( {{\chi }\o {\psi}}\)
 \label{topch}
 \er
Let us explicitly  
consider the $A_{r}^{(1)}$ model described by the 
Lagrangean density (\ref{lagran}) (with fields rescaled by $\varphi_i \rightarrow \b \varphi_i, 
\chi \rightarrow \b \chi, \psi \rightarrow \b \psi$, $\b^2 = -{{2\pi}\o {k}}$)
invariant under the following set of
discrete transformations, 
\br
\varphi_{j}^{\pr} = \varphi_{j} + {{2\pi (j-1) N}\o {\b_0 r}}, j=2, \cdots r,\quad 
\chi^{\pr} = e^{i\pi ({{N\o r} + s_2}) }\chi, \quad \psi^{\pr} = e^{i\pi ({{N\o r} + s_1}) }\psi
\er
where $s_1, s_2$ are both even or  odd integers and the following CP transformations (P: $x \rightarrow -x$)
\br
\varphi_{j}^{\pr \pr} = \varphi_{j}, j=2, \cdots r,\quad
\chi^{\pr \pr} = \psi, \quad \psi^{\pr \pr} =\chi
\er
 
The minimum of the potential ( for the choice $\eta =0$)
 corresponds to the following field configuration 
\br
\varphi_{j}^{(N)} ={{2\pi (j-1) N}\o {\b_0 r}}, \quad \theta^{(L)} ={{1}\o {2i \b_0}}ln \( {{\chi }\o {\psi }}\) = 
 {{\pi L}\o {\b_0}}, \quad \rho^{(0)} = 0 \quad j=2, \cdots r
 \label{vac}
\er
where $N, L$ are arbitrary integers, and the new fields $\theta$ and $\rho$ are defined as
\be
\psi = {1\o {\b_0}} e^{i \b_0 ({1\o 2} \varphi_2 - \theta )}\sinh (\b_0 \rho ), \quad 
\chi = {1\o {\b_0}} e^{i \b_0 ({1\o 2} \varphi_2 + \theta )}\sinh (\b_0 \rho )
\ee
In fact eqns, (\ref{vac}) also represent constant solutions of the eqns.
 of motion (\ref{13.20})-(\ref{1333}) which allows us to derive
the values of the topological charge (\ref{topch}): 
\br 
 Q_j = {1\o r}(j-1)(N_+-N_-),
 \quad  Q_{\theta} = -{{ \pi}\o {\b_0^2}} (L_+ - L_-)
 \nonu 
 \er

%

 
\sect{Zero Curvature and Equations of Motion}

The equations of motion for the NA Toda models are known to be of the form
\cite{lez-sav}
\be 
\bar \pa (B^{-1} \pa B) + [\hat {\eps_-}, B^{-1} \hat {\eps_+} B] =0, 
\quad \pa (\bar \pa B B^{-1} ) - [\hat {\eps_+}, B\hat {\eps_-} B^{-1}] =0
\label{eqmotion}
\ee
 The subsidiary constraint $J_{Y \cdot H^{(0)}} =
  Tr(B^{-1} \pa B Y\cdot H^{(0)})$ and $
  \bar J_{Y \cdot H^{(0)}} = 
  Tr(\bar \pa B B^{-1}Y \cdot H^{(0)} )=        0$ can be
 consistently imposed  since $[Y\cdot H^{(0)}, \hat {\eps_{\pm}}]=0$ as can be 
 obtained from
 (\ref{eqmotion}) by taking the trace with $Y.H^{(0)}$.  Solving those equations 
 for the nonlocal
 field $R$ yields, 
 \be
 \pa R = ({{Y\cdot \a_a}\o {Y^2}}) {{\psi \pa \chi }\o \Delta }e^{\Phi(\a_a)},
 \;\; \;\;\; 
\bar \pa R = ({{Y\cdot \a_a}\o {Y^2}}) {{\chi \bar \pa \psi }\o \Delta }e^{\Phi(\a_a)}
\label{bh}
\ee
The equations of motion for the fields $\psi, \chi $ and $\varphi_i, i=2,
\cdots , r$  obtained from (\ref{eqmotion}) after imposing the subsidiary 
constraints
(\ref{bh}) coincide precisely with the Euler-Lagrange equations derived from
 (\ref{action}).  Alternatively, (\ref{eqmotion}) admits a
zero curvature representation $\pa \bar A - \bar \pa A + [A, \bar A] =0$ where
\be
A= B \hat {\eps_-}  B^{-1} ,\quad  
\bar A= -\hat {\eps_+}   - \bar \pa B B^{-1} 
\label{zcc}
\ee
Whenever the constraints (\ref{bh}) are incorporated into $A$ and $\bar A$ in
(\ref{zcc}), equations (\ref{eqmotion}) yields the zero curvature
representation of the NA singular Toda models. 

We shall be considering 
$\hat \lie =  A_r^{(1)}$, $ Q= r\hat d + \sum_{i=2}^{r} 2 {{\lambda_i \cdot
H^{(0)}}\o {\a_i^2}}$,  ${\sum_{i=1}^{r} {Y_i H_i^{(0)}}} = 
2{{\lambda_1 \cdot H^{(0)}}\o {\a_1^2}}$,
 $ { \sum_{j=1}^{r}{X_i^j  H_j^{(0)}}} =
h_i^{(0)} = {{2 \a_i\cdot H^{(0)}}\o {\a_i^2}}$ and  $\hat {\eps_{\pm}} = 
\mu \(\sum_{i=2}^{r} E_{\pm \a_i}^{(0)} +
 E_{\mp (\a_2 + \cdots + \a_r)}^{(\pm 1)}\)$,  $\mu >0$.

Using the explicit parametrization of $B$ given in (\ref{63}),   we find, in
a systematic manner, the following form for $A$ and $\bar A$
 
\br
A  &=& \mu (  \sum _{i=2}^r e^{  -\sum _{j=2}^r K_{i,j}\varphi _{j}} E_{-\alpha 
_{i}}^{(0)}-\chi e^{-{1\o 2}R-2\varphi _{2}+\varphi _{3}}E_{-\alpha _{1}-\alpha 
_{2}}^{(0)}  \nonu \\
 &+&\psi e^{{1\o 2}R + \varphi_r -\eta} E_{\a_1 + \cdots \a_r}^{(-1)} + 
 (1+\psi \chi
e^{-\varphi_2})E_{\a_2 +\cdots + \a_r}^{(-1)} e^{\varphi_2 + \varphi_r -\eta
})
\label{a}
\er 
and 
\br 
\overline{A} =  \mu \( -{{\sum _{i=2}^r }}E_{\alpha 
_{i}}^{(0)}-E_{-\alpha _{2}-...-\alpha _{r}}^{(1)}\) -\( 
\overline{\partial }\chi -\chi  
\overline{\partial }\varphi _{2}+
({1\o {2\lambda_1^2}}-1){{\chi ^{2}\overline{\partial }\psi }\o \Delta } 
e^{-\varphi _{2}}\) e^{-{1\o 2}R} E_{-\alpha _{1}}^{(0)}  \nonu 
\er
\be 
 -   {{\overline{\partial }\psi }\o \Delta }
 e^{{1\o 2}R-\varphi _{2}}E_{\alpha _{1}}^{(0)}- 
\overline{\partial }\nu \hat {c}-\overline{\partial }\eta \hat {d} -{ 
{\sum_{i=2}^r }}\overline{\partial }\varphi _{i}h_{i}^{(0)}-{{\chi  
\overline{\partial }\psi}\o \Delta} e^{-\varphi _{2}}{ 
{\sum _{j=2}^r}}\left( \frac{r+1-j}{r}\right) h_{j}^{(0)}, 
\label{13.14} 
\ee 
 leading to the following equations of motion 
\begin{equation} 
\partial \overline{\partial }\eta =0, \quad 
 \partial \overline{\partial }\nu =\mu ^{2}e^{\varphi _{r}-\eta }(e^{\varphi 
_{2}}+{\psi }{\chi }),
\label{13.20} 
\end{equation}

\begin{equation} 
\partial \left( \frac{e^{-\varphi _{2}}\overline{\partial }{\psi } 
}{\Delta }\right) +\left( \frac{r+1}{2r}\right) \frac{{\psi } 
e^{-2\varphi _{2}}\partial {\chi }\overline{\partial }{ 
\psi }}{\Delta ^{2}}+\mu ^{2}e^{\varphi _{r}-\eta }{\psi }=0, 
\label{13.22} 
\end{equation} 
 
\begin{equation} 
\overline{\partial }\left( \frac{e^{-\varphi _{2}}\partial {\chi } 
}{\Delta }\right) +\left( \frac{r+1}{2r}\right) \frac{{\chi }%
e^{-2\varphi _{2}}\partial {\chi }\overline{\partial }{%
\psi }}{\Delta ^{2}}+\mu ^{2}e^{\varphi _{r}-\eta }{\chi }=0 
\label{13.23} 
\end{equation} 
 
\br
\pa \bar \pa \varphi_i &+& ({{r+1-i}\o {r}}){{\pa \chi \bar \pa \psi
e^{-\varphi_2}}\o {\Delta^2}} \nonu \\ 
&+& \mu^2 e^{\varphi_2 +\varphi_r -\eta }
\( 1+({{i-1}\o {r}})\psi \chi e^{-\varphi_2  }\)
 - \mu^2 e^{-\sum_{j=2}^{r}K_{ij}\varphi_j}  =0,  
 \label{1333}
 \er
$i=2, \cdots r$, where we  have normalized $\a^2 =2$.

The equations of motion (\ref{13.20})-(\ref{1333}) can be derived from the action
\br
S_{eff}&=&-\frac{k}{4\pi }\int d^{2}x\( {\sum _{i,j=2}^{r}}
K_{i,j}\partial \varphi _{i}\overline{\partial }\varphi
_{j}+\partial \nu \overline{\partial }\eta +
\partial \eta \overline{\partial }\nu    \right. \nonu \\
 &+& 
\left. \frac{2e^{-\varphi _{2}}\partial {\chi }\overline{\partial }
{\psi }}{\Delta }-2\mu ^{2}\( 
\sum _{i=2}^{r} e^{ {-\sum _{j=2}^{r} }K_{i,j}\varphi
_{j}} +e^{\varphi _{r}+\varphi _{2}-\eta }(1+{\psi }
{\chi }e^{-\varphi_2})\) \) .
\label{lagran}
\er
The fundamental Poisson bracket relation (FPR) can be derived for the 
$A_x = A - \bar A$ component of the two dimensional gauge connection.
 It relates the cannonical structure derived from the action (\ref{lagran}) 
 and a Lie algebraic structure by  the
 classical $r$ matrix.  Explicit construction  of the FPR is given in ref.  \cite{emilio}.

The model is  invariant under conformal transformations, i.e.
 $z \rightarrow f(z)$ e $\bar{z}
\rightarrow \bar{f}(\bar{z}),$ where ($f,\bar{f})$ 
are analytic functions and the fields transform as 
\br
&&{\chi }(z,\bar{z})\rightarrow \( \bar f^{\pr} \) ^{{1-r}\o 2}{\chi
}(f,\bar{f}), \quad  {\psi }(z,\bar{z})\rightarrow \(  f^{\pr} \) ^{{1-r}\o
2}{\psi }(f,\bar{f}), \nonu \\ 
 &&\eta (z,\bar {z})\rightarrow \eta
(f,\bar{f})  -r ln \( f^{\pr}(z) \bar f^{\pr}(\bar z)\) \nonu \\ 
&&\varphi_i(z, \bar z) \rightarrow \varphi_i(f,\bar{f}) - 
{{(i-1)(r-i+1)}\o 2}ln \( f^{\pr}(z) \bar f^{\pr}(\bar z)\), \quad i=2,
\cdots r\nonu \\
&&\nu (z,\bar {z})\rightarrow \nu (f,\bar{f})  +\d  ln \( f^{\pr}(z) \bar f^{\pr}(\bar z)\)
\label{confsym}
\er
where $\d $ is arbitrary.  The infinite dimensional conformal symmetry (\ref{confsym}) is generated by 
the chiral componentes of the improved stress energy momentum
tensor 
\br
T(z) &=&  -{{k}\o {4\pi }}( {1\o 2} K_{i,j} \pa \varphi_i  \pa \varphi_j + 
{{\pa \psi \pa \chi }\o {\Delta}}e^{-\varphi_2} + \pa \eta \pa \nu   \nonu \\
 &+& 
{{(r-1)\o 2}}\pa (
{{\psi \pa \chi }\o {\Delta }}e^{-\varphi_2} ) + \pa ^2 ( \sum_{i=2}^{r} \varphi_i + r \nu ) ) 
\er
and the same for $\bar T$ obtained by $\pa \rightarrow \bar \pa, \psi \rightarrow \chi $.

\sect{ Dressing Transformation, Vertex Operators and Degenerate Physical States }
\subsection{Dressing}
The conformal invariance may be broken by taking 
a particular solution of the first eqn.
 of
motion in (\ref{13.20}) namely $\eta = const $.  Equations 
(\ref{13.22})-(\ref{1333}) do
not 
depend on $\nu $ and can be solved independently.  The set of equations 
(\ref{13.22})-(\ref{1333}) with $\eta = const $ defines the affine NA Toda model 
and
their zero curvature representation  is given by (\ref{a}) and (\ref{13.14}) 
with
$\eta =const $.

A well established method for determining soliton solutions is provided by the
dressing  of a vacuum  into a non trivial solution by  gauge transformation
\cite{mira}.  
The zero curvature condition implies pure gauge connections,
 $A = -\pa T T^{-1}$ and 
$\bar A = -\bar \pa T T^{-1}$.  Suppose there exist a vacuum 
solution satisfying 
\be
\bar \partial B_{vac}B_{vac}^{-1}=\mu^2z\hat c, \quad 
B_{vac}\eps ^{-}B_{vac}^{-1}=\eps ^{-}  \label{8.13}
\ee
leading to 
\begin{equation}
A_{vac}=\eps ^{-}, \quad  
\overline{A}_{vac}=-\eps ^{+}-\mu ^2z\hat c.
 \label{8.15}
\end{equation}
where $[ \eps ^{+},\eps ^{-}]=\mu^2\hat c $.  
The solution for $A_{vac} = -\pa T_0 T_0^{-1}$ and 
$\bar A_{vac} = -\bar \pa T_0 T_0^{-1}$ is therefore given by 
\begin{equation}
T_{0} =\exp (-z\eps ^{-})\exp (\overline{z}\eps ^{+}).
\label{8.18}
\end{equation}
The dressing method is based on the assumption of the existence of two gauge
transformations generated by $\Theta^{\pm}$, mapping the vacuum into non trivial 
configuration,
i.e.
\begin{equation}
A=\Theta ^{\pm }A_{vac}(\Theta ^{\pm })^{-1}-\partial \Theta ^{\pm }(\Theta
^{\pm })^{-1}  \label{8.19}
\end{equation}
and 
\begin{equation}
\overline{A}=\Theta ^{\pm }\overline{A}_{vac}(\Theta ^{\pm })^{-1}-\overline{
\partial }\Theta ^{\pm }(\Theta ^{\pm })^{-1}  \label{8.20}
\end{equation} 
or 
\begin{equation}
B\eps ^{-}B^{-1}=\Theta ^{\pm }\eps ^{-}(\Theta ^{\pm })^{-1}-\partial \Theta
^{\pm }(\Theta ^{\pm })^{-1}  \label{8.21}
\end{equation}
and
\begin{equation}
-\eps ^{+}-\overline{\partial }BB^{-1}=\Theta ^{\pm }(-\eps ^{+}-
\mu^2z\hat c)(\Theta ^{\pm })^{-1}-\overline{\partial }
\Theta ^{\pm }(\Theta ^{\pm})^{-1}.  \label{8.22}
\end{equation}
As a  consequence we relate
\begin{equation}
\Theta ^{+}T_{0}=\Theta ^{-}T_{0}g,  \label{8.24}
\end{equation}
where $g$ is an arbitrary constant group element.
We suppose  that $\Theta^{\pm}$ are group elements of the form
\begin{equation} 
\Theta ^{-}=e^{t(0)}e^{t(-1)}e^{t(-2)} \cdots, \quad 
\Theta ^{+}=e^{v(0)}e^{v(1)}e^{v(2)} \cdots
  \label{8.26} 
\end{equation} 
where $t^{(-i)}, v^{(i)}$ are linear combinations of grade $(-i)$ and $(i)$
 generators
respectively ( $i=0,1, \cdots $).
In considering $\Theta^{-}$ the zero grade component of (\ref{8.21}) 
 admits solution
\be 
t(0) = e^{H(\bar z)}
\label{t0}
\ee
For $\Theta^+ $ the zero grade of  equation (\ref{8.22})   admits 
\be
e^{v(0)} = B e^{G(z) -\mu^2 z\bar z \hat c}
\label{v0}
\ee
where $H(\bar z), G(z) \in \lie_0^0$, i.e.  commute with both $\eps^{\pm}$.
From (\ref{8.22})  for $\Theta^{-}$ we get for the zero grade
\be
\bar \pa B B^{-1} = e^{H(\bar z)}[ t(-1), \eps ^{+}]e^{-H(\bar z)} 
+ \mu^2 z \hat c + (\bar \pa e^{H(\bar z)})e^{-H(\bar z)}
\label{t-1}
\ee
From  (\ref{8.21}) for $\Theta^+ $ we obtain  by comparing the zero grade
\be
0 = e^{v(0)} [ v(1), \eps ^- ]e^{-v(0)} - \pa e^{v(0)}e^{-v(0)}
\label{v1}
\ee
Inserting (\ref{v0}) in (\ref{v1}) we get
\be
B^{-1} \pa B - \mu^2 \bar z \hat c = e^{G(z)}[ v(1), \eps ^-]e^{-G(z)} 
- B^{-1}(\pa e^{G(z)})e^{-G(z)} B
\label{b1}
\ee
Multipling  (\ref{t-1}) and  (\ref{b1}) by an element of $\lie_0^0$ and taking the
trace we find that the choice $H(\bar z) = G(z) =0 $ implies the 
subsidiary conditions 
\be
Tr( B^{-1} \pa B \lie_0^0) = Tr (\bar  \pa B B^{-1}\lie_0^0) = 0.
\label{bh1}
\ee
and in eqns. (\ref{bh}).
From eqn. (\ref{8.24}) we find 
\begin{equation} 
...e^{-t(-2)}e^{-t(-1)}Be^{-\mu^2 \overline{z} 
z\hat c}e^{v(1)}e^{v(2)}...=T_{0}gT_{0}^{-1}.  \label{8.43} 
\end{equation}
hence, 
\begin{equation} 
<\lambda^{\pr}  
|B\exp (-\mu^2 z\overline{z}\hat c)|\lambda > =
< \lambda^{\pr}|T_{0}gT_{0}^{-1}|\lambda >.  \label{8.45} 
\end{equation}
where $| \lambda >$ and $<\lambda^{\pr}|$ are annihilated by  $\lie_>$ and 
$\lie_<$ respectively.
Explicit space time dependence for fields $\in \lie_0 / \lie_0^0$ is given by choosing
specific matrix elements, defining the so called tau functions,
\br
\tau_0 \equiv e^{\nu -  \mu^2 z \bar z} &=& < \lambda_0|T_{0}gT_{0}^{-1}|\lambda_0 >,
\nonu \\
\tau_R \equiv e^{{{rR}\o {r+1}}+\nu -  \mu^2 z \bar z} &=& < \lambda_1|T_{0}gT_{0}^{-1}|\lambda_1 >,
\nonu \\
\tau_j \equiv e^{\lambda_1 \cdot \lambda_j R+ \varphi_j +
\nu -  \mu^2 z \bar z} &=& < \lambda_j|T_{0}gT_{0}^{-1}|\lambda_j >,\;\;
j=2, \cdots r
\nonu \\
\tau_{\psi} \equiv e^{{1\o 2}{{r-1}\o {r+1}}R+\nu -  \mu^2 z \bar z}\psi &=&
 < \lambda_1|T_{0}gT_{0}^{-1}E_{-\a_1}^{(0)}|\lambda_1 >,
\nonu \\
\tau_{\chi} \equiv e^{{1\o 2}{{r-1}\o {r+1}}R+\nu -  \mu^2 z \bar z}\chi &=&
 < \lambda_1|E_{\a_1}^{(0)}T_{0}gT_{0}^{-1}|\lambda_1 >
 \label{tau}
\er
where for $SL(r+1)$, $\lambda_1 \cdot \lambda_j = {{r+1-j}\o {r+1}}$.

\subsection{Vertex Operators}

 Suppose we now write the constant group element $g$ in (\ref{8.24}) as 
 \begin{equation}
g=\exp [F(\gamma )],  \label{9.1}
\end{equation}
where $\gamma$ is complex parameter and choose $ F(\gamma )$ to be an eigenstate
of $\eps^{\pm}$, i.e.         
\begin{equation}
\lbrack \eps ^{\pm },F(\gamma )]=f^{\pm }(\gamma )F(\gamma ),  \label{9.2}
\end{equation}
where $f^{\pm}$ are specific functions of $ \gamma$.   
It therefore follows that,
\begin{equation}
T_{0}gT_{0}^{-1}=\exp \{\rho (\gamma )F(\gamma )\}  \label{9.6}
\end{equation}
where 
\begin{equation}
\rho =\exp \{-zf^{-}(\gamma )+\overline{z}f^{+}(\gamma )\},  \label{9.5}
\end{equation}
For more general cases where
\begin{equation}
g=\exp [F_{1}(\gamma _{1})]\exp [F_{2}(\gamma _{2})]...\exp [F_{N}(\gamma
_{N})],  \label{9.7}
\end{equation}
with
\begin{equation}
\lbrack \eps ^{\pm },F_{i}(\gamma _{i})]=f_{i}^{\pm }(\gamma
_{i})F_{i}(\gamma _{i}),  \label{9.8}
\end{equation}
we find
\be
T_{0}gT_{0}^{-1}=\exp [\rho _{1}(\gamma _{1})F_{1}(\gamma _{1})]\exp [\rho _{2}(\gamma
_{2})F_{2}(\gamma _{2})]...\exp [\rho _{N}(\gamma _{N})F_{N}(\gamma _{N})],
\label{9.9}
\end{equation}
where 
\begin{equation}
\rho _{i}(\gamma _{i})=\exp [-zf_{i}^{-}(\gamma _{i})+\overline{z}
f_{i}^{+}(\gamma _{i})].  \label{9.10}
\end{equation}
For the $A_r$  case, it is possible to show that $F^2 =0$ since $F(\gamma_i)$
is constructed in terms of vertex operators \cite{vertex}.  
In our specific model, the gradation interpolates between the 
homogeneous and the
principal gradations given by $Q=rD+{{\sum_{i=2}^r }}\frac{
2\lambda _{i}.H^{(0)}}{\alpha _{i}^{2}}$.  The constant grade $\pm 1$
 generators
are
\begin{equation}
\eps ^{+}=\mu \( {{\sum_{i=2}^r }}E_{\alpha
_{i}}^{(0)}+E_{-(\alpha _{2}+...+\alpha _{r})}^{(1)}\), \quad 
\eps^{-}=\mu \( {{\sum_{i=2}^r }}E_{-\alpha
_{i}}^{(0)}+E_{\alpha _{2}+...+\alpha _{r}}^{(-1)}\),
  \label{9.63}
\end{equation}
and their eigenstates are known to be \cite{aratyn}, \cite{vertex}
\begin{equation}
{F}_{1,j}^{\pm}(\gamma )=
{\sum_{n=-\infty }^{\infty }
}\gamma ^{-rn}{{\sum_{p=0}^{r-1}}}w^{\pm pj}\gamma
^{\mp p}E_{\pm (\alpha _{1}+...+\alpha _{(p+1)})}^{(n) },  \label{9.65}
\end{equation}

\br  
F_{a,j} &=& {{\hat c}\o {(w^a -1)}} 
+ \sum_{n\in Z} \gamma^{-rn} \sum_{i=1}^{r-1}
h_{i+1}^{(n)} \sum_{p=1}^{i} w^{a(i-p)} \nonu \\
&+& \sum_{b=1}^{r-1}
\sum_{n\in Z} w^{bj} \gamma^{-(b+nr)}\(\sum_{i=1}^{r-b} w^{a(i-1)}
E^{(n)}_{\a_{i+1} +\a_{i+2} + \cdots +\a_{i+b}}\right. \nonu \\  
& +& \left. \sum_{i=1}^{b} w^{a(i+r-1-b)}E^{(n+1)}_{-(\a _{i+1} + \a _{i+2}
+ \cdots + \a_{i+r-b})}\) 
\label{fal}   
\er
where $j=1,...,r, a=1,...,r-1$ and
$w=\exp \( \frac{2\pi i}{r}\)$.
 Their eigenvalues are obtained from
\begin{equation}
\lbrack \eps ^{\pm },{F}_{1,j}^{s }(\gamma )]=-s \mu w^{\mp j}\gamma^{\pm} 
{F}_{1,j}^{s }(\gamma ), \quad s= \pm  \label{9.71}
\end{equation}

\begin{equation}
\lbrack \eps ^{\pm },F_{a,j}(\gamma )]=\mu w^{\mp j}(w^{\pm a}-1)\gamma^{\pm}
F_{a,j}(\gamma ).  \label{9.75}
\end{equation}

Although there are $r(r+1)$ vertices we  can classify them into subsets 
such that
all vertices within a subset are equivalent in the sense that they 
all provide the
same solution by redefinition of  the spectral parameter. 
 In fact, such  equivalence is established directly from the form of the
 eigenstates.  It then follows 
 
\begin{equation} 
{F}_{1,r}^{\pm }(\gamma =\widetilde{\gamma }w^{-j})={F} 
_{1,j}^{\pm }(\gamma =\widetilde{\gamma }),  \label{14.25} 
\end{equation}

\begin{equation}
F_{a,r}(\gamma =\widetilde{\gamma }w^{-j})=F_{a,j}(\gamma =
\widetilde{\gamma }).  \label{14.27}
\end{equation}
For convenience we shall be considering solutions corresponding to one
representative of each subset, namely,
${F}_{1,r}^{\pm},$   $F_{d,r}$, $d=1, \cdots r-1$. 
 For  $\gamma ^{\ast }=\gamma ^{-1}$ one can find,
\begin{equation}
{{F}_{1,j}^{+}}^{\dagger }(\gamma )={F}_{1,j}^{-}(\gamma ), \quad 
F^{\dagger}_{a,j}(\gamma) = F_{r-a, (j-a) \; mod \; r}(\gamma)
\label{9.113}
\end{equation}

\subsection{Degenerate  Physical States - The Structure of Solutions}
In this section we shall classify  the soliton solutions, and discuss their
 properties in terms of the algebraic
structure of   the 
 $r+1$ vertex operators: $F^{\pm}_{1,r}, F_{d,r}, d=1, \cdots
 , r-1$.

 Appart from the vacuum solution, characterized by zero energy and momentum, the 
  basic soliton solutions are called 1-soliton ( 1-antisoliton ), the scattering of two
 1-solitons results in a 2-soliton  solution and so on.  There are also  bound states of two
 1-soliton solutions, whose basic property is that their
 energy   is smaller than the sum of energies corresponding to  two 1-soliton solutions.
 
 A solution constructed with a single  vertex $g= e^{F}$ where $F = F_{1,r}^{-}$ or $F_{1,r}^{+}$ 
 is a vacuum solution, since they contain only step operators and annihilate 
  the diagonal matrix elements in (\ref{tau}). Those matrix elements define the spectra (i.e., 
  mass, energy, momentum, electric charge).   
  In the first case the only nontrivial tau function is 
 \be
 \tau_{\chi} = a \exp [ -{{\mu z}\o {\g}} + \mu \g \bar z]
 \ee
 where $(a, \g )$ are complex constants.  The other tau functions vanish
 identically, i.e.
 \be
 \tau_0 = \tau_R = \tau_{\psi } = \tau_j =0
 \ee
 This implies that the field $\chi \neq 0 $ and $\psi =0$.  Since the stress
 energy momentum depend upon the product $\psi \chi $ and its derivatives, such
 solution corresponds to energy and momentum $E=P=0$.    Analogously for 
 $F_{1,r}^{+}$  where $\psi \rightarrow   \chi $ and similarly for the
sum   $\sum_{i=1}^{N} a_i F^{s}_{1,r}(\g_i), s= \pm 1$.  In other words, $N$ vertices
 $F_{1,r}^{-}$ or $F_{1,r}^{+}$ result in the physical vacuum (where $N$ is arbitrary).

 There are two basic 1-soliton solutions. They are provided by the constant group
 elements 
 \be
 g = \exp [c F_{d,r}(\g )]
 \label{f}
 \ee
 or 
 \be
 g = \exp [a F_{1,r}^-(\g_1 )]\exp [b F_{1,r}^+(\g_2 )]
 \label{ff}
 \ee
 The first provide electric charge $Q_{el} =0$, since the structure of $F_{d,r}$ in 
 eqns. (\ref{tau}) leads to $R=0$ ( $F_{d,r}$ does not contain $h_1^{(0)}$), 
  while in the second 
 $Q_{el} \neq 0$.  In this sense  one can say that $F_{d,r}$  are neutral vertices
 while both $F_{1,r}^-, F_{1,r}^+ $ are charged vertices.
 It will be shown in the next section that (\ref{f}) provides the solution of the
 $A_{r-1}^{(1)}$ abelian affine  Toda model.
 
 The solutions obtained from 
  \be
 g = \exp [c F_{d,r}(\g )]\Pi_{i=1}^{N} \exp [a_i F_{1,r}^{s}(\g_i)]
 \label{fff}
 \ee
 and 
 \be
 g = \exp [a F_{1,r}^-(\g_1 )]\exp [b F_{1,r}^+(\g_2 )]\exp [c
F_{1,r}^{s}(\g_3 )]  \label{ffff}
 \ee 
where $s = +1$ or $s = -1$ correspond to the same energy and momentum as those
 obtained from (\ref{f}) and (\ref{ff}) respectively. However their explicit 
 form ($\tau$-functions ) differ and may involve more
 parameters (see Sect. 5).

There are three basic  2-soliton solutions  characterized by
\br
 g = \exp [c_1 F_{d_1,r}(\g_1 )]\exp [c_2 F_{d_2,r}(\g_2 )]
 \label{2a}
 \er
 \be
  g = \exp [a F_{1,r}^-(\g_1 )]\exp [b F_{1,r}^+(\g_2 )] \exp [c F_{d,r}(\g_3 )] 
   \label{2b}
   \ee
   \be
  g = \exp [a F_{1,r}^-(\g_1 )]\exp [b F_{1,r}^+(\g_2 )]\exp [cF_{1,r}^-(\g_3 )]
   \exp [d F_{1,r}^+(\g_4 )]
   \label{2c}
   \ee
   It will be shown that the respective energies correspond to the sum of the 1-soliton
   individual energy. 
 
 The solution corresponding to (\ref{2a}) is the 2-soliton of the 
 $A_{r-1}^{(1)}$ abelian affine  Toda model.  If we now take 
  \br
 g = \exp [c_1 F_{d_1,r}(\g_1 )]\exp [c_2 F_{d_2,r}(\g_2 )]\Pi_{i=1}^{N} \exp [a_i
 F_{1,r}^{s} (\g_i)],
 \label{2ad}
 \er 
 $s= \pm 1$, the corresponding solutions will have the same energy and momentum as those
 obtained from (\ref{2a}) and 
  are also 2-soliton solutions.
  
  We have three basic bound states solutions.  As in the 2-soliton solutions they arise from 
  (\ref{2a}), (\ref{2b}) and (\ref{2c}), but the specific choice of the parameters $\g $ 
  is different.

\sect{1-Soliton Solutions}

\subsection{Neutral 1-soliton }
Our model (\ref{1.1}) has two basic independent  1-soliton solutions.  
The first one is obtained by taking the constant group element $g$ as
\begin{equation}
g=\exp [cF_{d,r}(\gamma )],  \label{15.1}
\end{equation}
where $F_{d,r}$ is given in (\ref{fal}), $c,\gamma $ are complex 
constants and  $d=1, \cdots , r-1$ describe the different species of the soliton.
The corresponding solution solution obtained from (\ref{tau}) is 
\be
\tau_0 =  \tau_R = 1+ e^{\zeta + H}, \quad 
 \tau_{\psi }= \tau_{\chi }=0; \quad \tau_{j}=
{1+w^{d(j-1)}e^{\zeta + H}},
\quad j=2,...,r 
\label{15.20}
\ee
where 
\be
H=2\mu (-1)^{{n}}\sin (\frac{\pi d}{r})\cosh(S) \( x-t\tanh(S) \), \quad n\in Z, \quad S\in R .
\label{aga}
\ee
$w=\exp( \frac{2\pi i}{r} )$ and $\zeta = \zeta_R + i \zeta_I$ are related to the specific parametrization
\begin{equation} 
\gamma =i(-1)^{{n}} w^{-{{d}\o 2}}e^{ S}, \quad c =(w^{d}-1) e^{\zeta}
\label{15.15} 
\end{equation} 
This choice of $\g $ is such that the solution will have real energy and momentum.  The parameter $S$ is the rapidity $|v| = \tanh
(S)$.  The real part of $\zeta = \zeta_R$ corresponds to shifts in space time  and its imaginary part $\zeta_I$ is in general
associated to the topological charge.

Since the  nonlocal field $R$ is constant it follows that $Q_{el }=0$.  It is in this sense that one can say that $F_{d,r}$ are
neutral vertices.
The topological charge is determined in terms of  non trivial zeros of the potential (see (\ref{vac})) by
$\varphi_j$, i.e.
\be
Q_j^{top} = {1\o {2\pi i}}\int_{-\infty}^{\infty} \pa_x \varphi_j dx = (j-1) Q_{mag}, \quad j=2, \cdots r
\ee
where $Q_{mag } =  \pm {1\o r} (d )|_{mod \;\;r}$

The mass is obtained from $\tau_0$ as 
 described in the appendix D  
 \be 
 M = {E \o {\cosh (S)}} = {P \o {\sinh S}} = {{2k\mu r }\o {\pi }}\sin ({{\pi d }\o {r}})
 \ee
 corresponding to the mass formula of the abelian affine Toda model  $A_{r-1}^{(1)}$ \cite{liao}.

An important remark  to be pointed out is the fact that degenerate solutions can be obtained by
replacing the constant group element $g$ in eqn. (\ref{15.1}) by 
\be
g =   \exp [ cF_{d,r}(\g ) ] \Pi_{i=1}^{N} \exp [a_i F^{s }_{1,r}(\g_i)]
\label{ggg}
\ee
with arbitrary $N$.  Since $F_{1,r}^{\pm}$ only contain step operators,
the resulting solution yields the same spectra of mass, electric and topological charges. 
For instance,   the group  element in (\ref{ggg}) with $ s=-1$ 
 produces the same tau functions as in (\ref{15.20}), the only exception being 
one nonvanishing  tau function, 
\be
\tau_{\chi } = \sum_{i=1}^{N} a_i \exp [-{{\mu z}\o {\g_i}}+ \mu \g_i \bar z ], \quad
\tau_{\psi} = 0
\ee
Consequently, the field $\chi $ is different from zero but $\psi =0$.  Since the energy momentum tensor 
depends explicitly on the product
$\psi \chi $ and on its derivatives, the solutions defined by $g$ given in (\ref{15.1}) and  in (\ref{ggg}) with $s=-1$ 
 are degenerate. 
 Similarly, one can check that
electric and topological charges remain unchanged.  The same is true with the solution obtained from $g$  in (\ref{ggg}) with $s=+1$
 interchanging $\psi $ by
$\chi $.


\subsection{Charged 1-Soliton Solution}


In a our previous paper \cite{backlund}, electrically charged topological 1-soliton  
 solution  have been constructed  using the method of
Backlund transformations.  Here, we will employ 
the dressing method with $g$ chosen as

\be
g=\exp \left[ a{F}_{1,r}^{-}(\gamma _{1})\right] \exp \left[ b{F 
}_{1,r}^{+}(\gamma _{2})\right]  \label{16.1} 
\end{equation} 
where ${F}_{1,r}^{\pm }$ 
 are given in (\ref{9.65})    and $a,b,
\gamma_1$ and $\gamma_2$ are complex constants.  
From (\ref{tau}) we    obtain the solution
\br
\tau_0 &=&1+e^{\zeta +2{F}}, \quad \tau_R = 
1 + (\Gamma_1)^{-r}e^{\zeta +2 F} \nonu \\
 \tau_j &=&1 + (\Gamma_1)^{j-r-1}e^{\zeta +2F}, \quad \tau_{\chi} = 
 \Gamma_2 e^{\theta +i G} e^{{{\zeta }\o 2}},
  \quad \tau_{\psi} = \Gamma_2 e^{-(\theta +i G)} e^{{{\zeta }\o 2}}
  \label{chargedsol}
\er
$j=2, \cdots r$, where 
 \begin{equation}
\Gamma _{1}=-\exp (-2i\overline{a}) = {{\g_1 }\o {\g_2}}, \quad \quad 
\Gamma _{2}=\sqrt{(1-\Gamma _{1})( ( \Gamma _{1}) ^{-r}-1) }.
 \label{16.26}
\end{equation}
\begin{equation}
F=\mu \cos (\overline{a})\cosh (\overline{b})
\(x -t\tanh (\overline{b}) \), \quad \quad 
G=\mu \sin (\overline{a})\sinh (\overline{b})
 \(t \coth (\overline{b}) - x  \)
 \label{fg}
\end{equation}
$\overline{a},  \overline{b}$ are real parameters, $\Theta, \zeta $ are complex parameters. 
 They correspond to a specific parametrization 
\begin{equation} 
\gamma _{1}=-\exp ( \overline{b}-i\overline{a}),  \quad \quad  
\gamma _{2}=\exp ( \overline{b}+i\overline{a})   \label{16.12} 
\end{equation}
\begin{equation} 
a=\sqrt{\frac{(1-\gamma _{1,2})\( 1-\left( \gamma _{1,2}\right) ^{r} 
\) }{\left( \gamma _{1,2}\right) ^{r}}}\exp \( \Theta + {1\o 2}\zeta \)  \label{16.17} 
\end{equation} 
and 
\begin{equation} 
b=\sqrt{\frac{(1-\gamma _{1,2})( 1-(\gamma _{1,2})^{r}) }{(\gamma 
_{1,2})^{r}}}\exp \( -\Theta + {1\o 2}\zeta\),
 \label{16.18} 
\end{equation} 
where $\g_{i,j} = {{\g_i}\o {\g_j}}$.  The choice of $(\g_1, \g_2)$ is such that ensures real energy and momentum.

The asymptotics of the field $R$  lead to the electric charge
\be
Q_{el} = -\frac{2kr}{\pi }\left[   sign( \cos( \bar{a}))
\( \frac{\pi }{2} sign (\bar{a})- \bar a \) -(n_{1}^{+}-n_{1}^{-})\frac{\pi }{r}\right], \quad n_1^{\pm} \in Z.
\label{16.33} 
\end{equation} 
and $Q_{el} =0$ if $\cos (\bar a) =0$.
The topological charge $Q_j^{top} = (j-1) Q_{mag}$ is given in terms 
of the asymtoptics of the fields $\varphi_j$ (\ref{vac}), 
\be
Q_j^{top} = {{ (j-1)}\o {r}}(n^+ - n_- )_{mod \;\; r}, \quad n^{\pm} \in Z.
\ee

Using again the argument explained in the appendix D, we evaluate the 
mass of the solution,   yielding 
\be
M = {E \o {\cosh (\bar b)}} = {P \o {\sinh (\bar b )}} = {{2k\mu r }\o {\pi }}|\cos (\bar a)|
 \ee
The same solution  was also found using first order differential
 equations  obtained from Backlund transformation
in ref. \cite{backlund}.
Following the discussion of Sect. 4.3, we shall show that solution associated with
\be
g = \exp [a F^{-}_{1,r}(\g_1 )]\exp [b F^{+}_{1,r}(\g_2 )]\exp [c F^{s }_{1,r}(\g_3 )]
\ee
will have the same energy and momentum  as the one derived above.  
The general solution can be calculated yielding,  for $s = -1$ 
\br
\tau_0 &=& 1+ ab {{(\g_{1,2})^r e^{F_a + F_b}}\o {(1- \g_{1,2})(1- (\g_{1,2})^r)}} + 
bc {{(\g_{3,2})^r e^{F_b + F_c}}\o {(1- \g_{3,2})(1- (\g_{3,2})^r)}} \nonu \\
\tau_R &=& 1+ ab  {{e^{F_a + F_b}}\o {(1- \g_{1,2})(1- (\g_{1,2})^r)}} + 
bc  {{e^{F_b + F_c}}\o {(1- \g_{3,2})(1- (\g_{3,2})^r)}} \nonu \\
\tau_j &=& 1+ ab {{(\g_{1,2})^{j-1} e^{F_a + F_b}}\o {(1- \g_{1,2})(1- (\g_{1,2})^r)}} + 
bc  {{(\g_{3,2})^{j-1}e^{F_b + F_c}}\o {(1- \g_{3,2})(1- (\g_{3,2})^r)}} \nonu \\
\tau_{\chi} &=& ae^{F_a} + c e^{F_c}, \quad \tau_{\psi} = be^{F_b} 
\er
where
\be
\g_1 = e^{\hat a + i \bar a}, \quad \g_2 = e^{\hat b + i \bar b}, \quad \g_3 = e^{\hat c + i \bar c}
\ee
$ \bar a, \hat a, \bar b, \hat b, \bar c, \hat c $ are real numbers and 
\br
F_a &= &  \mu t (\cos (\bar a) \sinh (\hat a) + i \sin (\bar a)\cosh (\hat a)) - 
  \mu x (\cos (\bar a) \cosh (\hat a) + i \sin (\bar a)\sinh (\hat a))\nonu \\
 F_c &= &F_a(\bar a \rightarrow \bar c,\hat a \rightarrow \hat c ), \quad
 F_b = -F_a(\bar a \rightarrow \bar b,\hat a \rightarrow \hat b )
 \nonu 
 \er
 Following appendix D, the energy and the momentum   for the case $
\lim \limits_{x \rightarrow +\infty}(F_a + F_b) =  \lim \limits_{x \rightarrow -\infty}(F_b + F_c) = +\infty$ 
 are given as, 
\br
E&=& -{{k\mu r}\o {\pi }} \( \cos (\bar a)\cosh (\hat a) + i \sin (\bar a)\sinh (\hat a) - 
\cos (\bar c)\cosh (\hat c) - i \sin (\bar c)\sinh (\hat c)\)\nonu \\
P&=&{{k\mu r}\o {\pi }} \( \cos (\bar a)\sinh (\hat a) + i \sin (\bar a)\cosh (\hat a) - 
\cos (\bar c)\sinh (\hat c) - i \sin (\bar c)\cosh (\hat c)\)
\label{tau00}
\er
The reality of the energy and momentum imply
\be
\sin (\bar a) \sinh (\hat a) = \sin (\bar c) \sinh (\hat c), \quad 
\sin (\bar a) \cosh (\hat a) = \sin (\bar c) \cosh (\hat c)
\ee
leading to 
\be
\hat a = \hat c , \quad \sin (\bar a) = \sin (\bar c), \quad \cos (\bar a) =\pm  \cos (\bar c).
\ee
The choice $\cos (\bar a) =  \cos (\bar c)$ gives zero mass configuration. The other choice 
results in  a perfect agreement with the mass formula already obtained  with $c=0$
\be
M = {{E}\o {\cosh (\hat a)}} = {{P}\o {\sinh (\hat a)}} = {{2k\mu r }\o {\pi }}|\cos (\bar a)|
\ee
Taking $\bar c = \pi sign (\bar a) - \bar a $  we get the same electric charge $Q_{el}$ in 
(\ref{16.33}).  The other cases with different asymptotic behaviour lead to
the same results.  The solution with $s= +1$ also gives the same conclusion.


\sect{Two Soliton Solutions and  Time Delays  }
\subsection{Two Soliton Solutions}
The 2-soliton solutions represent the scattering of two 1-soliton solutions. 
 These may be composed of neutral   or charged vertices given by products 
 of  exponentials associated to $F_{d,r}$ and $(F_{1,r}^+,
 F_{1,r}^-)$.  Combining the vertices we can obtain three different classes of 2-solitons:
 \begin{itemize}
\item  {\it neutral-neutral} (abelian) $(F_{d,r})\otimes (F_{d,r})$ 
\item  {\it neutral-charged}  $(F_{d,r})\otimes (F_{1,r}^+,F_{1,r}^-  )$
\item  {\it charged-charged}     $(F_{1,r}^+,F_{1,r}^-  )  \otimes (F_{1,r}^+,F_{1,r}^-  )$
\end{itemize}

The same vertices also provide  bound states solutions.  These differ 
by  the specific
parametrization of the spectral parameters. 
We will discuss, in this section the construction  of the {\it neutral-charged} 
case.  The final result of the other two cases is shown in the appendix C.
Finally we will discuss the spectrum of the three solutions.

Consider
\be
g = \exp[ a F_{1,r}^-(\g_1)]\exp[ b F_{1,r}^+(\g_2)]\exp[ c F_{d,r}(\g_3)]
\label{ttt}
\ee
where $a,b,c, \g_i $ are  arbitrary complex constants,
 $d=1, \cdots , r-1$ correspond to the {\it neutral} specie.
  Details of  the calculation  of the tau functions 
   (\ref{tau}) in terms of (\ref{ttt}) are given  in the appendix A.
 We therefore obtain, in general 
 \be
 \tau = \sum_{n_1, n_2, n_3=0}^{1} A_{n_1,n_2, n_3}^{\tau} \exp [ n_1 A_1(\g_1) + n_2 A_2(\g_2) +n_3 A_3(\g_3) ]
 \label{tt}
 \ee
 where $\tau = (\tau_0, \tau_R, \tau_j, \tau_{\psi}, \tau_{\chi})$,  $A_{n_1,n_2, n_3}^{\tau} $ 
 are certain functions of $\g_1,
 \g_2, \g_3, a,b,c$,  $A_1(\g_1) = -{{\mu z}\o {\g_1}} + \mu \g_1 \bar z$, 
$A_2(\g_2) = A_1(-\g_2)$ and $A_3(\g_3) = -{{\mu z}\o {\g_3}}(w^{-d} -1) + \mu \g_3 \bar z(w^{d} -1) $.
The tau functions given by (\ref{tt}) lead to solutions of the equations of motion (\ref{13.20})-(\ref{1333}), with $\eta =0$.
  However nothing 
is said about its solitonic
nature, which can be 2-soliton or a bound state.  The key point is to choose the spectral parameters 
in an appropriate way to ensure   the reality of the energy, momentum and electric 
charge (we emphasize that all
solutions, including the 1-soliton were obtained following the same principle).   
For the {\it neutral-charged} case (\ref{tt}) we have
\be
\tau_0 = 1+ A_{0,0,1}^{\tau_0} \exp {\( A_3(\g_3)\) } + A_{1,1,0}^{\tau_0} \exp {\( A_1(\g_1)+ A_2(\g_2) \) }+
 A_{1,1,1}^{\tau_0} \exp {\( \sum_{i=1}^{3}A_i(\g_i)\) }
 \label{tauzero}
 \ee
 and in general $(E,P)$ are proportional to linear combinations of 
 $\pa_x A_i(\g_i)$ and $\pa_t A_i(\g_i)$, since they depend upon the logarithmic derivatives of $\tau_0$. 
 The simplest choice is  $A_1(\g_1)+ A_2(\g_2)$ and $A_3(\g_3)$ to 
 be real.  One can see that this choice leads to $A_3
 \rightarrow H$ and $A_1+A_2 \rightarrow 2F$, where $H$ and $F$ are 
 the functions associated, respectively to the 1-soliton,
 neutral and charged, expressions 
 (\ref{aga}) and (\ref{fg}). 
  Implying that such 
 solution will have energy and momentum obtained by the
 combination of the corresponding 1-soliton values.  In this case, 
 the electric charge is already fixed to be real.  The other
 nontrivial choice leads us to the bound state solution discussed in 
 the next section.  The explicit 2-soliton solution for all
 three cases are presented in the appendix C.  The energy, 
  momentum and masses of the three solutions are obtained according to eqn.
 (\ref{17.83}) and  (\ref{17.84a})to be a surface term.  They correspond to 
 sums of quantities associated to the 1-soliton.

 The electric charge being evaluated as asymptotics of the non local
  field $R$ is also a surface term.  It can be verified that
 it is the sum of electric charges of 1-solitons.  Clearly, the 
 {\it neutral-neutral} is the abelian neutral 2-soliton and the
 {\it charged-charged} has two different contributions to $Q_{el}$.  
 The topological charge is also the sum of two 1-soliton
 quantities.  

\subsection{Time Delays}

The two soliton solutions describe the scattering of two
 1-soliton solutions. It is known that the effect
 of the scattering corresponds to a lateral displacement and associated 
 time delay \cite{kneipp}. 
  To make clear our discussion, consider 
  1-soliton solutions described by  the field  $\xi_a (x - x_{0a} - v_a t )$.
    Suppose that there
 exist a 2-soliton solution and let $\xi_{ab} (x_{0a},   x_{0b},  v_a, v_b )$ be 
 the associated field.  Suppose also that, by
 taking $x-v_at = const. $ we have 
 \be
 \lim \limits_{t \to \pm \infty }\xi_{ab} = K \xi_a (x-x_{0a}- v_at - x^{\pm}_a)
 \label{lim}
 \ee
 where $K$ is some complex constant.  It means that we are following the 1-soliton 
  $a$.  The lateral displacement is defined by 
 $ \Delta _a(x) = x_a^+ - x_a^- $ leading to the time delay 
 \be
 \Delta _a(t ) = - {{\Delta _a(x)}\o {v_a}}
 \ee

In order to analyse the time delays for our 2-soliton solutions,
  note that the model 
has more than one field.  
 Let us define a convenient combination of fields by
\be
\phi_j = \varphi_{j+1} - \varphi_{j} - {{R}\o {r+1}}, \quad j = 1, \cdots r
\ee
$\varphi_1 = \varphi_{r+1} =0$.  There are three classes of 2-soliton
solutions.   We will analyse the {\it neutral-charged} with details  and
present the conclusions for the other two cases 
(the {\it neutral-neutral} (abelian ) is already shown in the 
literature \cite{kneipp}).  The
expression for $e^{\phi_j}$ is 
\be
e^{\phi_j}= {{1+w^{dj}e^{\zeta_1 + H}+\Gamma_1^{-r+j}e^{\zeta_2 + 2F}+
w^{dj}\Gamma_1^{-r+j}\Gamma_{N,c}e^{\zeta_1+ \zeta_2}e^{  H+2F}}\o {
1+w^{d(j-1)}e^{\zeta_1 + H}+
\Gamma_1^{-r+j-1}e^{\zeta_2 + 
2F}+w^{d(j-1)}\Gamma_1^{-r+j-1}\Gamma_{N,c}e^{\zeta_1+ \zeta_2}e^{  H+2F}}}
\label{phi}
\ee
where
$H = 2 \mu (-1)^{n} \sin ({{\pi d}\o {r}}) \cosh(S)\(x- t\tanh (S) \)$ and 
$ F = \mu \cos(\bar a) \cosh (\bar b) \(x-t \tanh (\bar b) \)$
We take $\tanh (S)>0$,  $\tanh (S)- \tanh (\bar b)>0$ and follow the 
1-soliton associated to $H$: 
\be
x  - t \tanh (S) = const
\label{xt}
\ee
Substituting (\ref{xt}) in (\ref{phi}) we get for $\cos (\bar a)>0$
\be
\lim \limits_{t \to \infty }{\exp [\phi_j]} = 
\Gamma_1 \({{1+ e^{\zeta_1} w^{dj}\Gamma_{n,c}e^{H}}\o 
{1+ e^{\zeta_1} w^{d(j-1)}\Gamma_{N,c}e^{H}}}\)
\label{gamac}
\ee
\be
\lim \limits_{t \to -\infty }{\exp [\phi_j]} =  \({{1+ e^{\zeta_1} w^{dj}e^{H}}\o 
{1+ e^{\zeta_1} w^{d(j-1)}e^{H}}}\)
\label{gamad}
\ee
and for $\cos (\bar a)<0$, the limits are interchanged.  The expression (\ref{gamad})
 corresponds exactly to one soliton
solution, however expression (\ref{gamac}) is multiplied by a complex constant $\Gamma_1$ 
and has $H$ shifted by $ln (\Gamma_{N,c})$.  Thus, for 
$\cos (\bar a)>0$ $x_{H}^{-} = 0 $ and $x_{H}^{+} = 
- {{ln (\Gamma_{N,c})}\o {2 \mu (-1)^{n} \sin ({{\pi d}\o {r}}) \cosh (S)}}$.
 Then  
\be
\Delta_{H} (x)= {{-(-1)^{n} sign (\cos (\bar a))ln (\Gamma_{N,c})}\o {2 \mu
\cosh (S) \sin ({{\pi d}\o {r}})}} \label{deltah}
\ee
where 
\be
\Gamma_{N,c} = {{\cosh (S-\bar b) - (-1)^{n} \sin (\bar a + {{\pi d }\o {r}})}\o 
{{\cosh (S-\bar b) - (-1)^{n} \sin (\bar a - {{\pi d }\o {r}})}}}
\label{gammanc}
\ee
For $(-1)^n sign \( \cos (\bar a)\) = +1$, we have $\Gamma_{N,c} \in [ 0,1[ $ and hence 
   $ln (\Gamma_{N,c})<0$.  On the other hand, if $(-1)^n sign (\cos (\bar a)) = -1$, 
  $\Gamma_{N,c} \in [ 1,\infty [ $ and  
   $ln (\Gamma_{N,c})>0$.  Therefore, we can write both cases in a compact form
\be
\Delta_H(x) = {{|ln ( \Gamma_{N,c})|}\o {2 \mu \cosh (S) \sin ({{\pi d }\o
{r}})}} \geq 0  \ee
Following the arguments of ref. \cite{kneipp} and under the kinematics assumed
above, $\Delta_H (x) >0$ implies that the force  is attractive.  

Under the same conditions 
$\tanh (S)>0, \tanh (S)- \tanh (\bar b)>0$ but following the
1-soliton associated to $F$: 
\be
x - t\tanh (\bar b) = const 
\ee
we obtain
\be
\Delta_{F}(x) = -{{|ln ( \Gamma_{N,c})|}\o {2 \mu \cosh (\bar b) |\cos (\bar
a)|}} \label{latdispl}
\ee
Reversing the conditions 
$\tanh (\bar b)>0, \tanh (\bar b)- \tanh (S)>0$, $ \Delta_{H} \rightarrow - \Delta_{H}, 
\Delta_{F} \rightarrow - \Delta_{F}$.  
The time delay is for $\tanh (S)>0, \tanh (S)- \tanh (\bar b)>0$ 
\be
\Delta_H(t) = -{{|ln ( \Gamma_{N,c})|}\o {2 \mu \sinh (S) \sin ({{\pi d }\o
{r}})}}, \quad  \Delta_F(t) = {{|ln ( \Gamma_{N,c})|}\o {2 \mu \sinh (\bar b)
|\cos (\bar a)|}} \label{timedel}
\ee
Reversing the conditions for $(S, \bar b)$ leads to $ \Delta_{H}(t) \rightarrow - \Delta_{H}(t), 
\Delta_{F}(t) \rightarrow - \Delta_{F}(t)$.  The same analysis can be done for
the product $ (\psi \chi )$. 
  A subtle point is 
that if we follow the 1-soliton
associated to $H$, we get $\lim \limits_{t \to \pm \infty } \psi  \chi = 0 $ 
which corresponds 
to the abelian 1-soliton and does  not define the time delay. 
 Following the 1-soliton  associated to $F$, we get the 
corresponding $(\psi \chi )$ of the
charged 1-soliton solution in (\ref{chargedsol})
with $K = w^d$ and space displacement and time 
delay given by the same expressions above.  

The
analysis of the lateral displacements and time delays 
for the abelian {\it neutral-neutral} case is similar, replacing 
 $\Gamma_{N,c} $ by  $\Gamma_{N,N} $
interaction constant given by 
\be
\Gamma_{N,N} = {{\cosh (S_2 - S_1) - (-1)^{n_2-n_1} \cos ({{\pi (d_1 - d_2)}\o r})}\o 
{\cosh (S_2 - S_1) - (-1)^{n_2-n_1} \cos ({{\pi (d_1 + d_2)}\o r})}}
\ee
 where $d_1, d_2, S_1, S_2$ are defined in Appendix C. 
 The associated time delay  in the CMF is again negative leading to attractive forces. 
 
 Consider the solution given in (\ref{10.163}) corresponding to the 
  {\it charged-charged} case.   Taking  $v_{F_1} = \tanh  (\hat a)>0 ,  \tanh  (\hat a)> \tanh  (\hat f)$ 
and following the 1-soliton
associated to $F_1$:
\be
x- t \tanh (\hat a) = const
\ee
we get 
\be
\Delta_{F_1}(x) = -{{sign (\cos (\bar f))ln (\Gamma_{c,c} )}\o {2 \mu \cosh
(\hat a) \cos (\bar a)}} \ee
where
\be
\Gamma_{c,c}= {{\(\cosh (\hat a - \hat f ) - \cos (\bar a- \bar f )\) \(\cosh (r(\hat a - \hat f )) - 
\cos (r (\bar a- \bar f )) \)} \o {\(\cosh (\hat a - \hat f ) + \cos (\bar a+
\bar f )\) \(\cosh (r(\hat a - \hat f )) -  (-1)^{r} \cos (r (\bar a+ \bar f
)) \)}} \ee
Analysing the last two eqns. we conclude that  in general $\Delta_{F_1}(x)$ 
may be positive or negative.  Therefore  this solution represents 2-soliton solution
composed of  atractive or repulsive forces acting  between two charged 1-solitons.

\sect{Bound States}

The scattering of two basic 1-solitons gives rise to the 2-soliton and bound
state  solutions.  The most important diference between this last two kind of
solutions is  that the bound states have energy smaller than the corresponding
2-soliton solution.    Their difference defines the binding energy. 

 As explained in   the previous section we can obtain three classes of basic bound states:  {\it neutral-neutral} (abelian),
 {\it neutral-charged } and {\it charged-charged}.  We will discuss here the specific parametrization to the 
{\it neutral-charged} case.  The
 final results are shown in the appendix C for all the cases.  Consider the group element $g$ given by (\ref{ttt})  and the
 corresponding tau functions (\ref{tau}).  According to (\ref{17.83}),(\ref{17.84a})  and (\ref{tauzero}), 
 the energy and momentum are proportional to
 linear combinations of $\pa_x A_i(\g_i)$ and $\pa_t A_i(\g_i)$.  The most simple choice to obtain real $(E,P)$ is to take 
 $A_1(\g_1) + A_2(\g_2)$ and $A_3(\g_3)$ real giving rise to 2-soliton solutions.  However there is another alternative 
by taking $A_1(\g_1) + A_2(\g_2)$ and $A_3(\g_3)$ complex but the combinations 
$A_1(\g_1) + A_2(\g_2)\pm A_3(\g_3)$ to be real.
This leads to 
\br
A_3 \rightarrow 2\mu (-1)^{n} \sin ({{\pi d}\o {r}}) \cosh (S) [ (x- t \tanh
(S))\cos (q ) + i (-t +x \tanh (S)) \sin(q)] \nonu
\er
\br
A_1 + A_2 \rightarrow F_1 + F_2 
\er
 where
\br
F_1 = \mu \cosh (S) [ (x-t \tanh (S) )\cos (\bar a +\tilde a) + i (t - x \tanh (S) ) \sin (\bar a + \tilde a) ],\nonu
\er
\br
 F_2 = F_1 (\tilde a \rightarrow -\tilde a; i \rightarrow -i) 
 \er
  with the constraint
\be
\sin ({{\pi d }\o {r}}) \sin (q) + Sign [ (-1)^{n} \cos (q) \cos (\bar a )
\cos (\tilde a ) ] \cos (\bar a ) \sin (\tilde a ) = 0 
\ee
 The corresponding $(E,P,M)$ are given by 
 \be
 M = {{E}\o {\cosh (S)}} = {{P}\o {\sinh (S)}} ={{2kr\mu
}\o {\pi }} \( |\cos (\bar a ) \cos (\tilde a )| + |\cos (q)| \sin ({{\pi d }\o
{r}})\)
 \ee 
This defines the binding energy 
\br
\Delta E = {{2kr\mu
}\o {\pi }} \cosh (S) \( |\cos (\bar a )|( | \cos (\tilde a )| -1) + \sin
({{\pi d}\o {r}})(|\cos (q)|-1)\) \nonu
\er
The electric charge is also real and is given by 
\be
Q_{el} = -{{2kr}\o {\pi }}\( sign \( \cos (\bar a ) \cos (\tilde a)\) [ sign
(\bar a ) {\pi \o 2} - \bar a ] - (n_1^+ - n_1^-){\pi \o r}\)
\ee
for $\cos (\bar a ) \cos (\tilde a) \neq 0 $.  The $(E,P,M)$ for the other two bound states are 
\be
M = {{E}\o {\cosh (S)}} = {{P}\o {\sinh (S)}} ={{2kr\mu
}\o {\pi }} \( \sin({{\pi d_1}\o {r}})
|\cos(q_1)| + \sin({{\pi d_2}\o {r}}) |\cos(q_2)|\)
  \ee
for the {\it neutral-neutral} solution.  
For the  {\it charged-charged} case and under the conditions $sign (\cos (\bar
a + \tilde a)) = sign (\cos (\bar a - \tilde a)) $ and $sign (\cos (\bar
f + \tilde f)) = sign (\cos (\bar f - \tilde f)) $ 
 \be
M = {{E}\o {\cosh (\hat a)}} = {{P}\o {\sinh (\hat a)}} = {{2kr\mu
}\o {\pi }}\( |\cos (\bar a)
\cos (\tilde a )| + |\cos(\bar f ) \cos (\tilde f)| \)
\ee
Other conditions can be analysed  following the same procedure.
  The abelian case has zero electric charge  while within  the  
{\it charged-charged} case two different
contributions appear according to the combinations 
\br
 \cos (\bar a) \cos (\tilde a) >0 (<0) \;\;\;  {\rm and }\;\;\; 
 \cos (\bar f) \cos (\tilde f) >0 (<0).
\er

\sect{Outlook and Further Developments}

In this paper, the soliton solutions of the axial $A_r^{(1)}$ non abelian 
affine Toda model  were constructed and classified in terms of
vertex operators within the dressing formalism.  Those axial models are 
characterized by the presence of an antisymmetric term (torsion
term, see eqns. (\ref{axx1}) and (\ref{axx2})). 
 To each axial NA affine Toda model, a T-dual counterpart,
torsionless model were obtained systematically in terms of vector gauging 
of the two-loop WZW model \cite{wigner99}.  Their 1-soliton
solution for the $A_r^{(1)}$ case is presented in \cite{backlund}.

A subclass of T-selfdual non abelian affine Toda models have been established 
in connection with Kac-Moody algebras whose Dynkin diagram possess a ``$B_r$-tail like'' \cite{wigner99}, 
namely $B_r^{(1)}, A_{2r}^{(2)}$ and $D_{r+1}^{(2)}$ and coincide
 precisely with the models discussed by Fateev \cite{fat}.

A general study and classification of the soliton solutions of the torsionless T-dual 
and T-selfdual models in terms of dressing
formalism, $\tau$-functions and vertex operators are still to be completed and shall 
be reported elsewhere.

{\bf Acknowledgements} We  thank  L.A. Ferreira for many useful discussions. The  financial support of  
 CNPq, Fapesp and Unesp is gratefully acknowleged.

\sect{Appendix A}

Here we give an example of a typical calculation required to determine the tau functions. 
 Consider $g= \exp [aF_{1,r}^{-}(\g_1)]
\exp [bF_{1,r}^{+}(\g_2)]$.  Then since $T_0 = \exp (-z \eps_-) \exp (\bar z \eps_+)$ and 
\be 
[\eps^{\pm}, F^+_{1,r}(\g_2) ] =-\mu \g_2^{\pm } F^+_{1,r}(\g_2), \quad  
[\eps^{\pm}, F^-_{1,r}(\g_1) ] =\mu \g_1^{\pm } F^-_{1,r}(\g_1)
\ee
we get 
\be
T_0 g T_0^{-1} = \exp [aA_1(\g_1) F_{1,r}^{-}(\g_1)] \exp [bA_2(\g_2) F_{1,r}^{+}(\g_2)]
\ee
where
\be
A_1 (\g_1) = \(-{{\mu z }\o {\g_1}}+ \mu \bar z \g_1\), \quad A_2 (\g_2) = A_1 (\g_1\rightarrow -\g_2)
\ee
Let us now calculate
\br
\tau_0 &=& \langle \l_0 | T_0 g T_0^{-1} | \l_0 \rangle \nonu \\
&=&  \langle \l_0 | 1 + aA_1(\g_1) F_{1,r}^{-} + bA_2(\g_2) F_{1,r}^{+}(\g_2)
 + 
ab A_1(\g_1)A_2(\g_2)F_{1,r}^{-}(\g_1)F_{1,r}^{+}(\g_2)| \l_0 \rangle \nonu \\
\er
The last step cames from the fact that $F^{\pm}_{1,r}$ are associated to (nilpotent) vertex operators, 
i.e., their square inside the matrix
 element vanish.  Let 
\be
F_{1,r}^{\pm} (\g ) = \sum_{n=-\infty}^{\infty} \g^{-nr} \sum_{p=0}^{r-1} \g
^{\mp p} E^{(n)}_{\pm (\a_1 + \cdots + \a_{p+1})} \ee
We see that 
\be
\langle \l_0 | I| \l_0 \rangle =1, \quad \langle \l_0 | F_{1,r}^{\pm} (\g ) | \l_0 \rangle =0
\ee
where we have used the fact that $E^{(n)}_{ (\a_1 + \cdots + \a_{p+1})}| \l_0 \rangle =0, n\geq 0$.  Also
$\langle \l_0 | E^{(n)}_{ -(\a_1 + \cdots + \a_{p+1})}=0, n \leq 0$ and $[ E^{(n)}_{ -(\a_1 + \cdots + \a_{p+1})}, 
E^{(n)}_{ (\a_1 + \cdots + \a_{p+1})} ] = n\hat c \d_{m+n, 0} - \sum_{j=1}^{p+1} h_j^{(m+n)} $, leading to 
\be
\langle \l_0 | F_{1,r}^{-} (\g_1 ) F_{1,r}^{+} (\g_2 )| \l_0 \rangle = \sum_{n>0} n \g_{2,1}^{rn} \sum_{p=0}^{r-1} \g_{1,2}^{p} =
{{\g_{2,1}^r}\o {(1- \g_{2,1}^r)^2}} {{ (1- \g_{1,2}^r)}\o {(1- \g_{1,2})}}
\ee
where $\g_{ij} = {{\g_i}\o {\g_j}}$.

\sect{Appendix B}

In this appendix we want to show  that  the obtained tau functions do not depend upon the order in which the vertices are put in the
group element $g$.  To be more precise, let $g = \Pi_{i=1}^{N} e^{\a_i F_i(\g_i)}$, with $N$ arbitrary  and $F_i \in \{
F^{\pm}_{1,r}, F_{d,r}\}$ and define 
\br
\s_{j,k}g &=& \s_{j,k} \{ e^{\a_1 F_1(\g_1)}\cdots e^{\a_j F_j(\g_j)}\cdots e^{\a_k F_k(\g_k )}\cdots e^{\a_N F_N(\g_N)}\} \nonu \\
&=& \{ e^{\a_1 F_1(\g_1)}\cdots e^{\a_k F_k(\g_k)}\cdots e^{\a_j F_j(\g_j )}\cdots e^{\a_N F_N(\g_N)}\}
\er
for $j \neq k$, $j, k =1, \cdots , N$.  We want to show that $\tau (g) = \tau (\s _{jk} (g))$, where $\tau = (\tau_0, \tau_R,
\tau_{j}, \tau _{\psi}, \tau _{\chi} )$.  

Let us note that the vertex $F_i$ can be written as  $F_i = \sum_{n= -\infty}^{\infty} O(n) z^n$, where $z$ is some complex constant
and
 $O(n)$ are linear combinations of the generators 
with the $n$ dependence comming only from $(E_{\a}^{(n)}, h_i^{(n)})$.  Consider $\tau \neq (\tau_{\psi},\tau_{\chi})$, then 
\be
\tau (g) - \tau(\s_{j, j+1}(g)) =  \langle \l_l |\Pi_{i=1}^{N} e^{a_i A_i(\g_i) F_i(\g_i)}| \l_l \rangle - 
\langle \l_l |\s_{j, j+1} \Pi_{i=1}^{N} e^{a_i A_i(\g_i) F_i(\g_i)}| \l_l \rangle 
\label{b2}
\ee
where $l = 0, \cdots , r$ and $A_i(\g_i)$ are certain functions of $\g_i$.  Then we can expand the exponentials and to join the
corresponding terms.  For example, the last terms, 
\br
& &\langle \l_l |\Pi_{i=1}^{N} {a_i A_i(\g_i) F_i(\g_i)}| \l_l \rangle - 
\langle \l_l |\s_{j, j+1} \Pi_{i=1}^{N} a_i A_i(\g_i) F_i(\g_i)| \l_l \rangle \nonu \\
& =& a_1^2 \cdots a_n^2 A_1^2(\g_1) \cdots A_N^2(\g_N) 
\langle \l_l |F_1(\g_1) \cdots [F_j(\g_j),F_{j+1}(\g_{j+1})] \cdots F_N
(\g_N)| \l_l \rangle 
\label{b3}
\er
This expression corresponds to the sum of products of $N-1$ operators, since the commutator is 
equivalent to one operator.  But,
there are $N$ integer dummy variables (summed from $-\infty $ to $\infty$),  one for each vertex. 
 The standard algorithm  to
calculate the Kac-Moody matrix element is to reduce step by step the product of $N-1$ operators to 
one single operator that can be
evaluated.  At each step of this process one reduces the number of
unrestricted independent dummy  variables (i.e. $]-\infty , \infty [ $) by
one.  At the end of the process we have $N-1$ restricted (range $ \neq
]-\infty , \infty [ $) dummy variables  and one unrestricted.  From the
Kac-Moody commutation relations one can see that there are only terms 
proportional to ${z^{\pr}}^n$ or $n{z^{\pr}}^n$ for $n$
unrestricted ($z^{\pr}$ some complex constant)  and therefore according to the 
relation $\sum_{n = - \infty}^{\infty} {z^{\pr}}^n = 
\sum_{n = - \infty}^{\infty} n{z^{\pr}}^n = 0$ leads to 
 the vanishing of (\ref{b3}).  The same argument can
be used to show the vanishing of (\ref{b2}).  Also it is possible to do the same 
to $(\tau_{\psi},\tau_{\chi} )$.

\sect{Appendix C}

In this appendix we give the explicit solution of the three kinds of 2-solitons and bound states. 

a){\it  neutral-neutral}:  This solution is obtained from $g = \exp [ a F_{d_1, r}(\g_1)]\exp [ b F_{d_2, r}(\g_2)]$
\br
\tau_0 &=& 1+ e^{\zeta_1 + H_1} + e^{\zeta_2 + H_2} + \Gamma_{N,N} e^{\zeta_1 + \zeta_2}e^{H_1 + H_2}\nonu \\
\tau_j &=& 1+ w^{d_1(j-1)}e^{\zeta_1 + H_1} + w^{d_2(j-1)}e^{\zeta_2 + H_2} + 
\Gamma_{N,N} w^{(d_1 + d_2)(j-1)}e^{\zeta_1 + \zeta_2}e^{H_1 + H_2}, j=2,
\cdots r\nonu \\ \tau_R &=& \tau_0, \quad \tau_{\psi} = \tau_{\chi} = 0
\er
where $d_1, d_2 = 1, \cdots r-1$, $\zeta_1, \zeta_2 \in C  $, $w = e^{{{2\pi i }\o r}}$.  For the 2-soliton
\br
H_k&=& 2 \mu (-1)^{n_k} \sin ({{\pi d_k}\o {r}}) \cosh (S_k) \( x -t \tanh
(S_k)\) , \quad k=1,2 \nonu \\ \Gamma_{N,N}&=& {{\cosh (S_1 -S_2) -
(-1)^{(n_2-n_1)} \cos ({{\pi }\o {r}} (d_1 - d_2))}\o  {\cosh (S_1 -S_2) -
(-1)^{(n_2-n_1)} \cos ({{\pi }\o {r}} (d_1 + d_2))}}, \quad S_k \in R, \quad
n_k \in Z \er For the bound state, 
\br
H_k &=& 2 \mu (-1)^{n_k} \sin ({{\pi d_k}\o {r}}) \cosh (S) ( (x-t \tanh (S))
\cos (q_k) \nonu \\ &+& i (-t +x \tanh (S)) \sin (q_k) ), \nonu \\
\Gamma_{N,N}^{(b)} 
&=&\Gamma_{N,N}^{(2-s)} (S_2-S_1 \rightarrow i(q_2-q_1)), \quad S ,q_2,q_1 
\in R \er
with the constraint
\be
\sin ({{\pi d_1 }\o {r}}) \sin (q_1) + Sign [ ( -1)^{(n_1+n_2)} \cos (q_1)
\cos (q_2) ] \sin ({{\pi d_2 }\o {r}}) \sin (q_2) =0
\ee

b) {\it Neutral-Charged}:This solution is obtained from
\be 
g = \exp [ a F_{1, r}^{-}(\g_1)]\exp [ b F_{1, r}^{+}(\g_2)]\exp [ c F_{d, r}(\g_3)],
\ee
\br
\tau_0 &=& 1 + e^{\zeta_1 +H}+ e^{F_1+F_2}e^{\zeta_2} + \Gamma_{N,c} e^{\zeta_1 + \zeta_2}e^{H + F_1 +F_2}\nonu \\
\tau_R &=& 1 +  e^{\zeta_1 +H}+ \Gamma_{1}^{-r}e^{F_1+F_2}e^{\zeta_2} + \Gamma_{1}^{-r}\Gamma_{N,c} 
e^{\zeta_1 + \zeta_2}e^{H + F_1 +F_2}\nonu \\
\tau_j &=& 1+ w^{d(j-1)}e^{\zeta_1 + H} +
\Gamma_{1}^{-r+j-1}e^{F_1+F_2}e^{\zeta_2} 
+w^{d(j-1)}\Gamma_{1}^{-r+j-1}\Gamma_{N,c}e^{\zeta_1 + \zeta_2 }e^{H + F_1+
F_2} \nonu \\  
\tau_{\chi}&=& = \Gamma_2 e^{\theta + {1\o 2}\zeta_2} e^{F_1}
(1 + \Gamma_3 e^{\zeta_1 + H}), \nonu \\  \tau_{\psi} &=& \Gamma_2 e^{-\theta +
{1\o 2}\zeta_2} e^{F_2} (1 + \Gamma_4 e^{\zeta_1 + H}) 
\er
 where $d= 1, \cdots
,r-1, \theta ,\zeta_1, \zeta_2 \in C$, $\Gamma_1, \Gamma_2 $ are defined in (\ref{16.26}).  

For the 2-soliton. 
\be
F_1 = F + iG, F_2 = F-iG, 
\ee
\br
F&=&  \mu \cos (\bar a) \cosh (\bar b) \( x - t \tanh (\bar b) \) \nonu \\
G&=&  \mu \sin (\bar a) \cosh (\bar b) \( t  - x \tanh (\bar b) \) \nonu \\
H&=& 2 \mu (-1)^{n} \sin ({{\pi d}\o {r}})\cosh (S )\( x - t \tanh (S) \)
\er
\be
\Gamma_3^{(2-s)} = w^d {{1-i (-1)^n e^{-(S-\bar b)} e^{-i \bar a} w^{-{1\o 2}d}
}\o  {1-i (-1)^n e^{-(S-\bar b)} e^{-i \bar a} w^{{1\o 2}d}}}, \quad
\Gamma_4^{(2-s)} =\Gamma_3^{(2-s)} \((S-\bar b ) \rightarrow -(S-\bar b)\)
\ee
\be
\Gamma_{N,c}^{(2-s)} \equiv w^{-d} \Gamma_3^{(2-s)}\Gamma_4^{(2-s)} = {{\cosh (S-\bar b) - (-1)^{n} 
\sin (\bar a +{{\pi d }\o {r}} ) } \o {\cosh (S-\bar b) - (-1)^{n} 
\sin (\bar a -{{\pi d }\o {r}} ) }}, \quad S, \bar b, \bar a \in R, n \in Z
\ee
For the bound state
\be 
\Gamma_3^{(b)} = \Gamma_3^{(2-s)}(S-\bar b \rightarrow i(\tilde a + q )), \quad \Gamma_4^{(b)} =\Gamma_3^{(b)} 
( i(\tilde a + q )\rightarrow - i(\tilde a + q ) )
\ee
\be
\Gamma_{N,c}^{(b)} = \Gamma_{N,c}^{(2-s)} (S-\bar b \rightarrow i(\tilde a + q ) ), \quad \tilde a, q \in R
\ee
\br
F_1 &=&  \mu \cosh (S) [ (x-t \tanh (S)) \cos (\bar a + \tilde a) +
 i (t - x \tanh (S)) \sin (\bar a + \tilde a ) ],
 \nonu \\
H &=& 2 \mu (-1)^{n} \sin ({{\pi d }\o {r}}) \cosh (S) [ (x-t \tanh (S))
\cos (q) +  i (-t +x \tanh (S))\sin (q) ] \nonu \\
\er
$F_2 = F_1 (\tilde a \rightarrow -\tilde a; i \rightarrow -i ) $ with the constraint
\be
 \sin ({{\pi d }\o {r}})\sin (q) + sign [ (-1)^n \cos (q) \cos (\bar a) \cos
(\tilde a ) ] \cos (\bar a ) \sin (\tilde a  )= 0 
 \ee
 
c) {\it Charged-Charged}:This solution is obtained from 
\be
g = \exp [ a F_{1, r}^{-}(\g_a)]\exp [ b F_{1, r}^{+}(\g_b)]\exp [ c F_{1, r}^-(\g_f)]\exp [ g F_{1, r}^{+}(\g_g)],
\ee
\br
\tau_0 &=& 1 + e^{\zeta_1 +F_1 + F_3}+ e^{F_2+F_4}e^{\zeta_2} + K_1 e^{ F_1 +F_4}+ K_2 e^{ F_2 +F_3}
+ \Gamma_{c,c} e^{ F_1 +F_2+ F_3 +F_4}e^{\zeta_1 + \zeta_2} \nonu \\
\tau_R &=& 1 +   \Gamma_{1}^{-r}e^{F_1+F_3}e^{\zeta_1} + \Gamma_{2}^{-r} 
e^{\zeta_2 }e^{ F_2 +F_4}+ \Gamma_{3}^{-r}K_1e^{ F_1 +F_4} \nonu \\
&+& \Gamma_{4}^{-r}K_2 e^{ F_2 +F_3} +
(\Gamma_1\Gamma_2)^{-r} \Gamma_{c,c}e^{ F_1 +F_2 + F_3+ F_4} e^{\zeta_1 + \zeta_2} \nonu \\
\tau_j &=& \tau_R (\Gamma_i^{-r} \rightarrow \Gamma_i ^{j-r-1}), i=2, \cdots
r \nonu \\ 
 \tau_{\chi}&=&   e^{\theta_1 + {1\o 2}\zeta_1}\Omega_1  e^{F_1}
+  e^{\theta_2 + {1\o 2}\zeta_2}\Omega_2 e^{F_2} +   A_1 e^{F_1 + F_2 + F_4 }
+ A_2 e^{F_1 + F_2 + F_3 }, \nonu \\  \tau_{\psi} &=& e^{-\theta_1 + {1\o
2}\zeta_1}\Omega_1  e^{F_3} +  e^{-\theta_2 + {1\o 2}\zeta_2}\Omega_2 e^{F_4}
+   A_3 e^{F_1 + F_3 + F_4 } + A_4 e^{F_2 + F_3 + F_4 }  \label{10.163}
\er
where
\br
\Gamma_1 &=& \g_{a,b}, \;\; \Gamma_2 = \g_{f,g} \;\; \Gamma_3 = \g_{a,g} \;\; \Gamma_4 = \g_{f,b}\;\; 
\Gamma_5 = \g_{b,g} \nonu \\
\Omega_i &=&\sqrt{(1-\Gamma_i)(\Gamma_i^{-r} -1)}, \;\; i=1,2, \quad 
\Omega_j^{-1} =\sqrt{(1-\Gamma_j)(\Gamma_j^{-r} -1)}, \;\; j=3,4,\nonu \\
K_1 &=& e^{\theta_1 - \theta_2 } e^{{1\o 2}(\zeta_1 + \zeta_2) } \Omega_1
\Omega_2 \Omega_3^2, \;\;  K_2 = e^{\theta_2 - \theta_1 } e^{{1\o 2}(\zeta_1 +
\zeta_2 )}\Omega_1 \Omega_2 \Omega_4^2 
\er
where $\zeta_1, \zeta_2, \theta_1, \theta_2 \in C$, and 
\br
A_1 &=& e^{\theta_1 + {1\o 2} \zeta_1 + \zeta_2} (\Gamma_2 - \Gamma_3)
(\Gamma_2^{r} - \Gamma_3^{r})(\Gamma_2 \Gamma_3)^{-r} \Omega_1 \Omega_3^2\nonu
\\ A_2 &=& e^{\theta_2 + {1\o 2} \zeta_2 + \zeta_1} (\Gamma_4 - \Gamma_1)
(\Gamma_4^{r} - \Gamma_1^{r})(\Gamma_1 \Gamma_4)^{-r} \Omega_2 \Omega_4^2\nonu
\\ A_3 &=& e^{-\theta_2 + {1\o 2} \zeta_2 + \zeta_1} (\Gamma_3 - \Gamma_1)
(\Gamma_3^{r} - \Gamma_1^{r})(\Gamma_1 \Gamma_3)^{-r}  \Omega_2
\Omega_3^2\nonu \\ 
A_4 &=& e^{-\theta_1 + {1\o 2} \zeta_1+ \zeta_2} (\Gamma_2 -
\Gamma_4) (\Gamma_2^{r} - \Gamma_4^{r})(\Gamma_2 \Gamma_4)^{-r}  \Omega_1
\Omega_4^2\nonu \\ \er
For the 2-soliton,
$F_1 = F+i G, \quad F_3 = F-i G$
\br
F&=&  \mu \cos (\bar a) \cosh (\hat a) (x-t \tanh (\hat a)), \nonu \\
G&=&  \mu \sin (\bar a) \cosh (\hat a) (t-x \tanh (\hat a)), \nonu \\
 (F_2, F_4) &=& (F_1, F_3)(\bar a \rightarrow \bar f, 
\hat a \rightarrow \hat f )
\er
\be
\g_a = -e^{\hat a -i \bar a}, \;\; \g_b = e^{\hat a +i \bar a}, \;\; \g_f = - e^{\hat f - i \bar f}, \;\; 
\g_g = e^{\hat f + i \bar f}
\ee
\be
\Gamma_{c,c}^{(2-s)} = {{(\cosh(\hat a- \hat f) - \cos(\bar a - \bar f))\( \cosh (r(\hat a - \hat f)) - 
\cos (r(\bar a - \bar  f))\)}\o {(\cosh(\hat a- \hat f) + \cos(\bar a + \bar
f))\( \cosh (r(\hat a - \hat f)) -  (-1)^{r}\cos (r(\bar a + \bar  f))\)}}
\ee
where $\bar a, \hat a, \bar f, \hat f \in R$.

For the bound state
\br
F_1 &=&  \mu \cosh (\hat a) \( (x-t \tanh (\hat a))\cos (\bar a+ \tilde a) +
i (t-x \tanh (\hat a))\sin (\bar a + \tilde a)\)\nonu \\
F_3 &=& F_1 (\tilde a \rightarrow -\tilde a; i \rightarrow -i), \quad F_2 = 
F_1 (\tilde a \rightarrow \tilde f ,
\bar a \rightarrow  \bar f ), \quad F_4 = F_1 (\tilde a \rightarrow -\tilde f,
\bar a \rightarrow  \bar f;i \rightarrow -i )\nonu \\
\g_a &=& -e^{\hat a -i (\bar a + \tilde a)}, \quad \g_b = e^{\hat a +i (\bar a - \tilde a)}, \quad
\g_f = -e^{\hat a -i (\bar f + \tilde f)}, \quad
\g_g = e^{\hat a +i (\bar f - \tilde f)}\nonu \\
\Gamma_{c,c}^{(b)} &=& \Gamma_{c,c}^{(2-s)} (\hat a - \hat f \rightarrow
i(\tilde a - \tilde f) ),\quad   \hat a, \bar a, \tilde a,  \bar f,
\tilde f \in R, 
\er
with the constraint
\be
\cos (\bar a) \sin (\tilde a) + sign [ \cos (\bar a) \cos (\tilde a) \cos
(\bar f) \cos (\tilde f) ] \cos(\bar f) \sin (\tilde f) =0
\ee

\sect{Appendix D}

In order to  evaluate the energy and momentum of the system it is
 convenient to
 use the
procedure of ref. \cite{aratnp}.
The canonical energy-momentum tensor defined from the action  (\ref{lagran}) does
 not satisfy the
traceless condition. Instead, 
\begin{equation} 
T^{\alpha }{}_{\alpha }=2V,  \label{17.61} 
\end{equation} 
where $V$ is the potential appearing in the lagrangean density (\ref{lagran}).  Using
the equations of motion (\ref{13.20})-(\ref{1333}), we find, 
\begin{equation} 
T^{\alpha }{}_{\alpha }=4\partial _{\rho }\partial ^{\rho }f=4\partial  
\overline{\partial }f,  \label{17.62} 
\end{equation} 
where 
\begin{equation} 
f=-\frac{k}{4\pi }\left\{ {{\sum_{i=2}^r }}\varphi _{i}+ 
\frac{r(r-1)}{2(r+1)}R+r\nu \right\} .  \label{17.63} 
\end{equation}
and therefore define a traceless conformal energy-momentun tensor 
\be
\Theta _{cat}^{\mu \nu }=T^{\mu \nu }+
4 \left( \partial ^{\mu }\partial ^{\nu }-g^{\mu \nu }\partial _{\rho }
\partial ^{\rho }\right) f  \nonu
\ee

In ref. \cite{aratnp} it is argued that the total energy and total momentum are given
by
\begin{equation} 
\int_{-\infty }^{\infty }dxT^{00} =E =
-4\left( \partial _{x}f\right) |_{x\rightarrow 
-\infty }^{x\rightarrow \infty }  \label{17.68} 
\end{equation}

\begin{equation} 
\int_{-\infty }^{\infty }dxT^{01}=P  = 
4\left( \partial _{t}f\right) |_{x\rightarrow 
-\infty }^{x\rightarrow \infty }
.  \label{17.69} 
\end{equation} 
It follows that 
$ e^{-{{4\pi }\o{k}}f} = \tau_0 \tau_2 \cdots \tau_r e^{r \mu^2 z \bar z}$ 
 and therefore for the energy momentum we have
  
\begin{equation} 
E=-\left( -\frac{k}{\pi }\right) \left[ \frac{ 
\partial _{x}\tau_{0}}{\tau_{0}}+\frac{\partial
_{x}\tau_{2}}{\tau_{2}}+...+\frac{  \partial
_{x}\tau_{r}}{\tau_{r}}+\frac{\partial _{x}\(e^{r\mu ^{2}z\overline{z}}\)}{ 
e^{r\mu ^{2}z\overline{z}}}\right] _{x\rightarrow -\infty }^{x\rightarrow 
\infty }   \label{17.75}
\end{equation} 
and   
\begin{equation} 
P=\left( -\frac{k}{\pi }\right) \left[ \frac{ 
\partial _{t}\tau_{0}}{\tau_{0}}+\frac{\partial
_{t}\tau_{2}}{\tau_{2}}+...+\frac{  \partial
_{t}\tau_{r}}{\tau_{r}}+\frac{\partial _{t}\(e^{r\mu ^{2}z\overline{z}}\)}{ 
e^{r\mu ^{2}z\overline{z}}}\right] _{x\rightarrow -\infty }^{x\rightarrow 
\infty }.   \label{17.76}
\end{equation}
the last term in eqn (\ref{17.76})  does not  contributes, while in (\ref{17.75})
contributes with an infinite value (zero point energy), which will be regularized.

From the explicit form of the solution,  it is easy to check directly 
 that  each term
in (\ref{17.75}) and in (\ref{17.76}) yields the same asymptotic contribution 
and hence
 
\begin{equation} 
E=\frac{kr}{\pi }\left( \frac{\partial _{x}\tau_{0}}{ 
\tau_{0}}\right) _{x\rightarrow -\infty }^{x\rightarrow \infty } 
\label{17.83}  \end{equation}  
 
\begin{equation} 
P=-\frac{kr}{\pi }\left( \frac{\partial _{t}\tau_{0}}{ 
\tau_{0}}\right) _{x\rightarrow -\infty }^{x\rightarrow \infty }. 
\label{17.84a} 
\end{equation} 

\end{document}